\pdfoutput=1
\documentclass[aj]{emulateapj}

\DeclareRobustCommand\Will[1]  {#1}
\DeclareRobustCommand\Bob[1]   {#1}
\DeclareRobustCommand\Gary[1]  {#1}
\DeclareRobustCommand\Nick[1]  {#1}
\DeclareRobustCommand\Alert[1] {#1}
\DeclareRobustCommand\Final[1] {#1}
\DeclareRobustCommand\FFinal[1]{#1}
\DeclareRobustCommand\Info[1]{}
\DeclareRobustCommand\Note[1]{}
\DeclareRobustCommand\PNote[2][0.97\linewidth]{}
\DeclareRobustCommand\MNote[2][10cm]{} 

\providecommand\rmxaa{RMAA}     
\newcommand\thC{\ensuremath{\theta^1\,}Ori~C}
\newcommand{\kms}{\ensuremath{\mathrm{km\ s}^{-1}}}
\newcommand\htwo{H$_{\rm 2}$}
\newcommand\Halpha{H$\alpha$}
\newcommand\Hbeta{H$\beta$}

\newcommand\ori{\thC}
\newcommand\Htwo{H$_{2}$}
\newcommand\chbeta{\ensuremath{c_\mathrm{H\beta}}}
\newcommand\UpDown[2]{^{\smash{\scriptscriptstyle\mathrm{#1}}}_{\smash{\scriptscriptstyle\mathrm{#2}}}}
\newcommand\cBal{\ensuremath{c\UpDown{H\alpha}{H\beta}}}
\newcommand\cVLA{\ensuremath{c\UpDown{VLA}{H\beta}}}
\newcommand\cdif{\ensuremath{c\UpDown{DIF}{H\beta}}}
\newcounter{ionstage}
\renewcommand{\ion}[2]{\setcounter{ionstage}{#2}%
  \ensuremath{\mathrm{#1\,\scriptstyle\Roman{ionstage}}}}

\newcommand\Vsun{$V_{\sun}$}
\newcommand\Vomc{$V_{\rm OMC}$}
\newcommand\Vcluster{$V_{\rm Cluster}$}
\newcommand\Vori{$V_{\theta ^{1}\rm Ori~C}$}
\newcommand\Vtrap{$V_{\rm Trap}$}
\newcommand\VHplus{$V_{\rm Veil-H^{+}}$}
\newcommand\VA{$V_{\rm H~I-A}$}
\newcommand\VB{$V_{\rm H~I-B}$}
\newcommand\Vmif{$V_{\rm MIF}$}
\newcommand\Vmil{$V_{\rm MIL}$}
\newcommand\Vform{V$_{\rm H_{2}CO}$}
\newcommand\form{H$_{2}$CO}

\newcommand{\ition}[2]{\setcounter{ionstage}{#2}%
  \ensuremath{\mathit{#1\,\scriptstyle\Roman{ionstage}}}}
\newcommand\wind{_\mathrm{w}}
\newcommand\elec{_\mathrm{e}}
\newcommand\smyr{\ensuremath{M_\odot~\mathrm{yr}^{-1}}}
\newcommand\pcc{\ensuremath{\mathrm{cm^{-3}}}}
\newcommand\sbunits{\ensuremath{\mathrm{photons\ cm^2\ s^{-1}\ sr^{-1}}}}
\newcommand\fluxunits{\ensuremath{\mathrm{cm^{-2}\ s^{-1}}}} 
\newcommand\hii{\ion{H}{2}}
\newcommand\hi{\ion{H}{1}}
\newcommand\eff{_\mathrm{eff}}
\newcommand\oiii{[\ion{O}{3}]}
\newcommand\nii{[\ion{N}{2}]}
\newcommand\sii{[\ion{S}{2}]}
\newcommand\hei{[\ion{He}{1}]}

\slugcomment{Submitted to the AJ}

\shorttitle{Three Dimensional Structure of the Inner Orion Nebula}

\shortauthors{\uppercase{O'Dell et al.}}
\shorttitle{\uppercase{Three Dimensional Structure of the Inner Orion Nebula}}

\begin{document}

\title{The Three Dimensional Dynamic Structure of the Inner Orion Nebula\footnotemark[1]}

\footnotetext[1]{Based on observations with the NASA/ESA Hubble Space Telescope,
obtained at the Space Telescope Science Institute, which is operated by
the Association of Universities for Research in Astronomy, Inc., under
NASA Contract No. NAS 5-26555.}

\author{
  C. R. O'Dell,\altaffilmark{2}
  W. J. Henney,\altaffilmark{3}
  N. P. Abel,\altaffilmark{4}
  \Gary{G. J.} Ferland,\altaffilmark{5}
  \& S. J. Arthur\altaffilmark{3}
}

\altaffiltext{2}{Department of Physics and Astronomy, Vanderbilt University, Box 1807-B, Nashville, TN 37235}
\altaffiltext{3}{Centro de Radioastronom\'{\i}a y Astrof\'{\i}sica, Universidad Nacional Aut\'onoma de M\'exico, Apartado Postal 3-72,
58090 Morelia, Michaoac\'an, M\'exico}
\altaffiltext{4}{Department of Mathematics and Physics, College of Applied Science, University of Cincinnati, Cincinnati, OH 45221}
\altaffiltext{5}{Department of Physics and Astronomy, University of Kentucky, Lexington, KY 40506}

\email{cr.odell@vanderbilt.edu}

\begin{abstract}
  The three dimensional structure of the brightest part of the Orion
  Nebula is assessed in the light of published and newly established
  data.  We find that the widely accepted model of a concave blister
  of ionized material needs to be altered in the southwest direction
  from the Trapezium, \Will{where we find that the Orion-S feature is
    a separate cloud of very optically thick molecules within the body
    of ionized gas, which is} probably the location of the multiple
  embedded sources that produce the optical and molecular outflows
  that define the Orion-S star formation region. \Will{Evidence for
    this cloud comes from the presence of \form\ lines in absorption
    in the radio continuum and discrepancies in the extinction derived
    from radio-optical and optical only emission.}  \Bob{We present
    \Alert{an equilibrium} Cloudy model of the Orion-S cloud, \Alert{which successfully reproduces many observed properties o}f this feature,
    including the presence of \Alert{gas-phase} \form\ in absorption.}
  We also report the
  \Will{discovery of an open-sided shell of \oiii{} surrounding the
    Trapezium stars, revealed through emission line ratio images and
    the onset of radiation shadows beyond some proplyds. We show that
    the observed properties of the shell are consistent with it being
    a stationary structure, produced by shock interactions between the
    ambient nebular gas and the high-velocity wind from \ori.}
  \Alert{We examine the implications of the recently published}
  evidence for a large blueshifted velocity of \ori\ with respect to
  the Orion Molecular Cloud\Will{, which} could mean that this star
  has only recently begun to photoionize the Orion Nebula. \Alert{We
    show that current observations of the Nebula do not rule out such
    a possibility, so long as the ionization front has propagated into
    a pre-existing low-density region. In addition, a young age for
    the Nebula would help explain the presence of nearby proplyds with
    a short mass-loss timescale to photoablation.}

\end{abstract}

 

\begin{deluxetable*}{llcl}
\tablecaption{Heliocentric Velocities of the Major Orion Nebula Components}
\tablewidth{0pt}
\tablehead{
\colhead{Designation} & \colhead{Components}  & \colhead{\Vsun (\kms)} & \colhead{Source}}
\startdata 
\Vomc & Molecules & \(25.8\pm1.7\) & \citet{ode08b}\\
\Vmif & [\ion{O}{1}],[\ion{S}{2}] & \(25.5\pm1\) & \citet{ode01b}\\
\Vmil & [\ion{O}{2}],\oiii{},\nii{},[\ion{Cl}{3}],H$^{+}$,He$^{+}$ & 18.0$\pm$1.4 & \cite{ode01b}\\
\Vtrap & $\theta ^{1}\,$Ori~A,$\theta ^{1}\,$Ori~B,$\theta ^{1}\,$Ori~D & \(24\pm3\) & \citet{abt91}\\
\Vcluster & Lower Mass Stars & 25.8$\pm$1.0 & \citet{fur08}\\
\Vori & \ori & 13 & \citet{sta08}\\
\Alert{\(V_\mathrm{Arc}\)} & \Alert{Big Arc} & \(9.2\pm1\)  & This study.\\
\VHplus & S$^{++}$,P$^{++}$,\nii{},\ion{He}{1},[\ion{O}{2}] & \(3\pm2\) & \citet{abe06}\\
\VA    &  \hi{}           &  \(23.4\pm0.01\) & \citet{abe06}\\
\VB    &  \hi{}           &   \(19.3\pm0.03\) & \citet{abe06}\\
\Vform & \form\ Absorption & \(28.6\pm1.6\) & \citet{man93}\\
\enddata
\tablecomments{~Local Standard of Rest velocities can be obtained by
  subtracting 18.1 \kms\ from \Vsun\ values.
}
\end{deluxetable*}

\section{Introduction}

The brightest portion of NGC 1976, the Orion Nebula, is commonly
called the Huygens region (after \Will{Christiaan Huygens}, who
published the first drawing \Will{of the nebula in 1659}) and its form
in three dimensions was the subject of many early papers (reviewed in
\citealp{ode01a}. The presently accepted basic model of a photoionized
thin layer of gas flowing off the side of the Orion Molecular Cloud
(OMC) facing us was invoked to explain the progressive blueshift of
the emission lines with respect to the OMC \citep{zuc73,bal74}.  The
Bright Bar along the southeast border of the Huygens region is a
location where the main ionization front (MIF) is tilted almost along
the line of sight to the observer. The other portions of the MIF have
been mapped in three dimensions (3-D) by \citet{wen95} by a method
that assumes all the ionizing radiation arises from the \Alert{O7Vp}
star \ori, and then calculates the distance to the MIF that would
satisfy the known conditions of gas density and observed \Halpha\
surface brightness, this being a broader application of a method first
presented in \citet{bal91}. Although this 3-D map has the limitation
that it becomes progressively less accurate as one moves away from the
line of sight towards \ori, it did establish that \ori\ lies
about\footnote{Throughout this paper we will adopt a distance of 440
  pc to the Orion Nebula, a value derived \citep{ode08a} using the
  results of what are currently thought to be the best independent
  determinations. Earlier papers that used different assumed distances
  will have their results scaled to this distance.} 0.2~pc in front of
the MIF, it confirmed the structure in the Bright Bar region, and
showed that the nebula was otherwise a concave surface with a large
bump to the southwest of \ori\ in the Orion-S star formation region.
The surface brightness near \ori\ can be explained as emission by a
constant density layer of about 0.1~pc \Will{thickness}
\citep{pog92}. This is only a reference number since the emissivity
must be much higher near the MIF and drop as the square of the
density. In a nebula where there is free expansion of the photoionized
gas and there is a single dominant ionizing star, a concave shape of
the MIF is the natural result, with bumps, valleys, and ridges
reflecting underlying conditions of the host molecular cloud. The
ionization front is expected to be closer to the ionizing star where
the molecular cloud density is higher and further \Will{away} where
the underlying density is lower. The ionized gas density drops rapidly
away from the MIF, but it is not clear what the density is in the
immediate vicinity of \ori\ since it is expected that its intense high
velocity wind would create a hot low density cavity around it. This
cavity has not been detected directly except for observations of
stand-off shocks in front of the proplyds closest to \ori\
\citep{bal98,bal00}.

In addition to the rich Orion Nebula Cluster (ONC) centered on the
bright Trapezium stars, there are two star formation centers imbedded
in the OMC. The first is associated with the deeply imbedded (0.2 pc,
\citealp{doi04}) BN-KL infrared and radio sources to the northwest of
\ori\ and the second in the Orion-S region.
 
With the discovery of 21 cm absorption lines in the radio continuum
spectrum \citep{van89} it was recognized that there was a Veil of
neutral material on the observer's side of the nebula and subsequent
absorption line spectroscopy \citep{abe04b,abe06} established its
physical characteristics and approximate location of about one parsec
on the observer's side (henceforth foreground) of \ori.  \Will{Recent
  detailed reviews have covered the ONC \citep{mue08} and the nebula
  plus obscured star-formation regions (\citealp{ode08b}, and see also
  \citealp{ode01a,ode01b}).} 
 
The Huygens region occupies the northeast corner of a much larger
structure called the extended Orion Nebula (EON, \citealp{gud08}). It
is known that there is a systematic flow of ionized material into the EON \citep{ode01a,hen05}, but the lower surface brightness has limited the number of investigations of this region \citep{sub01}. However, the EON is the location of two X-ray bright regions of hot plasma \citep{gud08}. High resolution optical \citep{hen07} and infrared \citep{meg06} images of the EON are now available. Although these images show many interesting large-scale features, there is neither the detail nor abundance of stars of the Huygens region. 

In this paper we will present the most relevant information about the ONC and the Orion Nebula, then integrate this into a modified picture of the Orion Nebula's 3-D structure and its history. In \S\ 2 we present the most useful information, then give the results of where this information leads in \S\ 3.

\section{Background Information}
\subsection{Emission Line Images of the Orion Nebula}
The Huygens region has been the subject of numerous imaging
studies. Arguably the most useful groundbased study is that of
\citet{pog92}, which utilized a Fabry-Perot system to isolate emission
from the \Halpha, \Hbeta, \oiii{} 5007~\AA, \nii{} 6583~\AA{} and
6548~\AA, \sii{} 6716~\AA\ and 6731~\AA, and \hei{} 6678~\AA{}
lines. \citet{ode96} presented a mosaic of Hubble Space Telescope
(HST) WFPC2 images at the superior resolution of the HST. The
particular advantage of the HST images is that the WFPC2 images allow
clear discrimination of not only the isolated \oiii{} 5007 \AA\ line,
but also both the \Halpha\ 6563 \AA\ line and the nearby \nii{} 6583
\AA\ line in addition to the filters being narrow enough that they
provide a good isolation of the emission lines against the strong
scattered light continuum that primarily arises from \Will{dust
  grains} in the dense region immediately beyond the MIF
\citep{ode99}.  A later HST survey \citep{hen07} with the ACS camera
covered a wider field of view with pixels one half the angular size of
those in the WFPC2, but, the filters used don't allow a clear
delineation of the important emission lines \citep{ode04}. The Huygens
region has been mapped with the VLA at about 1.7\arcsec\ resolution
\citep{ode00} at the extinction free 20.5 cm continuum. By comparing
the surface brightness calibrated \citep{ode99} images and the radio
images, it was possible to derive a map of the optical extinction
across the Huygens region and to generate extinction-corrected
versions of the HST WFPC2 emission line images \citep{ode00}.  In this
study we employ the several forms of the WFPC2 images, all processed
in part using the IRAF package\footnote{IRAF is distributed by the
  National Optical Astronomy Observatories, which is operated by the
  Association of Universities for Research in Astronomy, Inc.\ under
  cooperative agreement with the National Science foundation.}. We see
in Figure 1 that the Orion-S region to the southwest of the Trapezium
resembles the Bright Bar region in being enhanced in low ionization
\nii{} emission and being much brighter than adjacent areas. These
characteristics are consistent with the brightest part of the central
Huygens region being both closer to \ori\ and also being an inclined
face of ionized gas. This picture is consistent with the results of
the 3-D modeling of \citet{wen95}.

\begin{figure*}
\includegraphics[width=\linewidth]{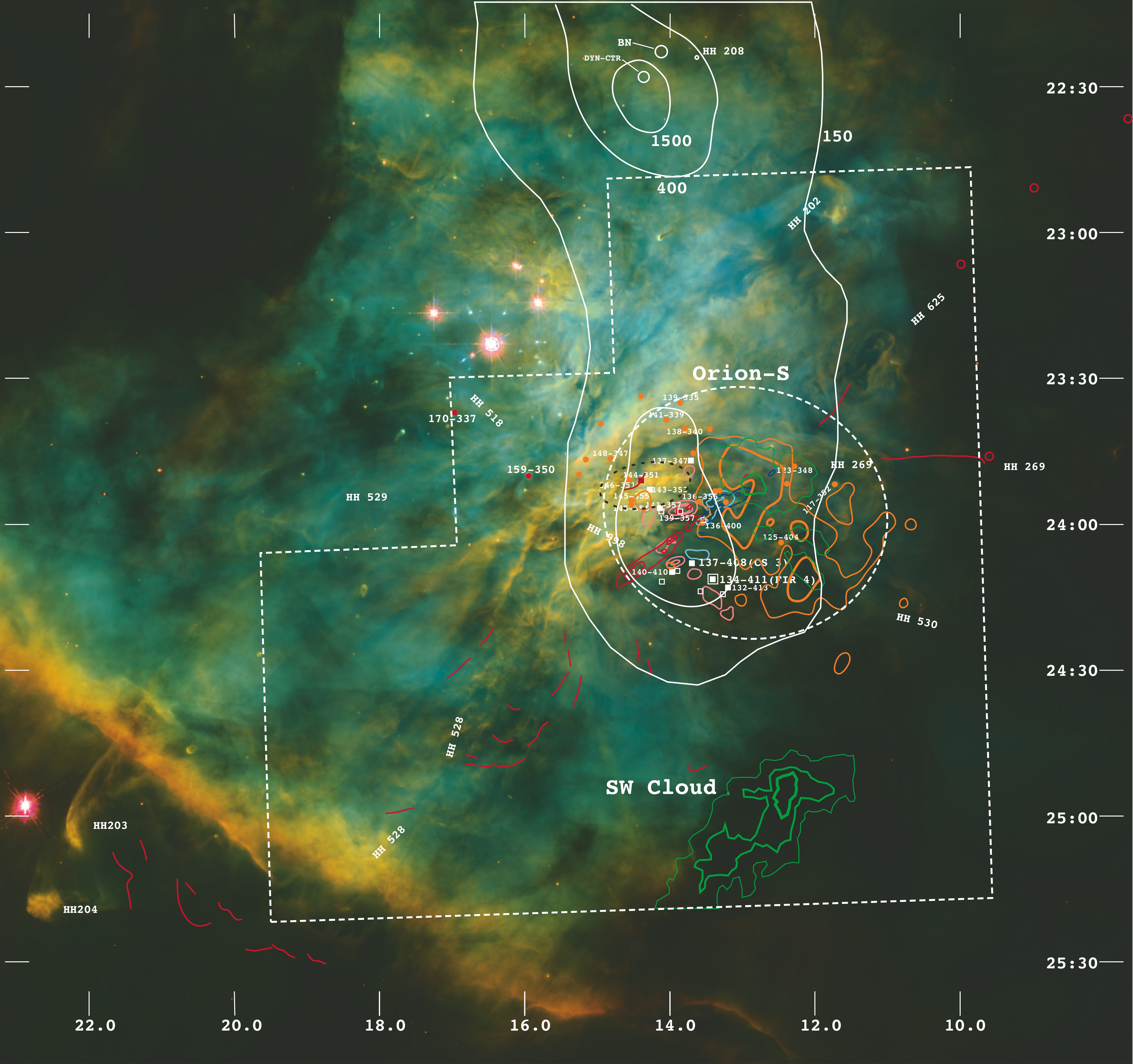}
\caption{\Alert{This \(233\arcsec \times219\arcsec\) image is composed
    of a mosaic of WFPC2 images \citep{ode96} with F502N (\oiii{}) as
    blue, F656N (\Halpha) as green, and F658N (\nii{}) as red.  North
    is up and the labels along the edge depict the Right Ascension
    beyond 5:30:00 and the Declination south of \(-5\):20:00
    (2000). Major outflow features are labeled in addition to objects
    discussed in the text. Throughout this paper a position based
    designation is used \citep{ode94} except for large individual features such as Herbig Haro objects. The white dashed ellipse encloses the smaller features collectively discussed in the text as the Orion-S feature. Red lines and circles represent
    strong infrared \Htwo\ features \citep{kai00,sta02} not associated
    with the BN-KL deeply embedded sources. The strong dark red/blue
    contoured lines indicate CO outflow \citep{zap05} and the pastel
    colored pink/light-blue contoured lines indicate SiO outflow
    \citep{zap06}. The orange contours are \form\ absorption features
    \citep{man93} at the level of \(-80 \mathrm{K}~\kms\) (heavier)
    and \(-50 \mathrm{K}~\kms\) (lighter).  The green contours of
    increasing thickness depict regions of the difference of
    extinction (\cdif\ as defined in \S\ 2.4) of 0.1, 0.2, and
    0.3. The sharp boundary of the more southerly \cdif\ excess
    feature is the result of reaching the edge of the WFPC2 field of
    view. The region most likely to contain the sources of the high
    velocity optical outflows is shown as a dark dashed ellipse
    \citep{ode08a}.  Open squares indicate the positions of H$_{2}$O
    maser sources \citep{gau98}. Point sources within the dashed
    outline are coded by the shortest wavelength of their detection,
    with filled white squares indicating the positions of radio only
    visible sources \citep{zap04a,zap04b,zap05}, red squares the
    positions of sources seen only in the mid infrared
    \citep{smi04,rob05}, and filled orange circles positions of stars
    in the near infrared catalog of \citet{hil00}.  The white contours
    show the 350 $\mu$m emission in this area \citep{hou04} in units
    of Jy per 12\arcsec\ beam. The irregular dashed white line
    indicates the field where \chbeta\ was determined both by the
    radio/optical and optical line ratio methods.  }
\label{master}
}
\end{figure*}


\subsection{The Orion Nebula's Veil}

Although the optical extinction of the Veil has been recognized for
some time, it has been established only recently \citep{ode92,ode02}
that most of this extinction arises from the near side of the ionized
zone and not within it.  The detailed analysis of \citet{abe05,abe06}
established that the physical conditions in the two 21 cm \hi{}
absorption velocity components of the Veil are rather different, the
energy density in their Component A being dominated by the magnetic
field measured from the Zeeman effect.  
\Will{The elements C, S, Mg, and Si are ionized in the Veil by the FUV
  (5--13.6~eV) radiation that penetrates it, even though the Veil is
  optically thick to EUV radiation.}  There must be a secondary
\Will{hydrogen} ionization front associated with the Veil and on the
far side (away from the observer and closer to \ori) of the Veil.
\cite{abe06} identify emission lines probably arising from the Veil's
ionization front.

The radial velocities of the various emission and absorption line components along the line of sight through the Trapezium are summarized in Table 1.

\subsection{\form\ is Seen in Absorption in Orion-S}
\label{sec:form-seen-absorption}

An early study \citep{joh83} at 16\arcsec\ resolution detected \form\
in absorption against the radio continuum with the location identified
as Orion-S, a region with multiple known molecular emission lines
\citep{ode08b} and well defined bipolar molecular outflows in CO
\citep{zap05} and SiO \citep{zap06}.  A higher resolution
(\(5.1\arcsec \times 7.6\arcsec\)) study \citep{man93} confirmed the
presence of this absorption feature and the results are shown in
Figure 1 and Table 1. In Figure 1 we see that the \form\ absorption is distinct
from the bipolar outflows, the strongest infrared and radio sources,
and the sources of the high velocity optical outflows
\citep{ode08a}. 
\citet{man93} point out that the presence of \form\ in
absorption means that the cloud containing it must lie in front of the
ionized gas, a position also taken in more general form by
\citet{joh83} and \citet{wil01}.  Since \form\ can only exist in a
cold dense gas that is optically thick to FUV (and therefore also EUV)
radiation, there will be a corresponding high extinction at visual
wavelengths.  Therefore, we have looked for optical extinction
associated with this feature with the results described in the next
section.

\subsection{Anomalies in the Extinction in the Orion-S Region}

The optical appearance of the Huygens region is strongly affected by
extinction occurring within the Veil. The clearest example is the Dark
Bay feature to the east-northeast of the Trapezium. This extinction
generally decreases away from the Dark Bay as the line of sight is
moved to the southwest. The extinction has been derived in several
fashions. In slit spectroscopy, the common approach has been to compare
the flux ratios of the strongest Balmer series lines with the ratios
expected from theoretical predictions calculated for the local
conditions (primarily the electron temperature).  A good example of
this is the study of \citet{bal91}, who obtained sample spectra along
a well-defined east-west path beginning about 30\arcsec\ west of \ori\
and derived the extinction from comparing Paschen lines with
H$\gamma$.  The highest spatial resolution study is that of
\cite{ode03}, where calibrated HST WFPC2 \Halpha\ and \Hbeta\ images
from a single WFPC2 pointing to the southwest of the Trapezium were
compared with theory. Pixel by pixel extinction corrections were
obtained as part of a study of electron temperature fluctuations. The
widest field of view high resolution determination of the extinction
was that of \citep{ode00} who compared the \Halpha\ surface brightness
of a mosaic of gaussian blurred HST WFPC2 images with VLA 20 cm images
obtained at 1.7\arcsec\ resolution. In all these studies similar
extinction curves were used \citep{cos70,car89}. The derived
extinction is commonly expressed as \chbeta, the logarithm of the
extinction at 4861~\AA, the wavelength of \Hbeta.  In this work we
will refer to the \chbeta\ value derived from the comparison of the
radio continuum and \Halpha\ as \cVLA and to the value derived from
comparison of WFPC2 Balmer \Halpha\ and \Hbeta\ images as \cBal, while
the difference of the two will be \(\cdif = \cVLA -\cBal\).

Although a comparison of the results for \cVLA\ and \cBal\ for nine well-sampled smaller regions showed that there was a good agreement of the results of the two methods for finding \chbeta\ (ratio of 0.93$\pm$0.19,\citep{ode00}, there is reason to expect that the two methods should not agree exactly. 

The first reason would be the effect of scattering of the optical lines by dust in the dense photon dominated region (PDR) that lies immediately beyond the MIF.  This scattering is what produces a broad redshifted component in the high resolution spectra of the intrinsically narrower heavy ion emission lines and it must also be present in the Balmer lines. This is why the continuum of the Orion Nebula is much stronger than expected from simply atomic processes \citep{bal91}, in effect, the Orion Nebula is also a bright reflection nebula. One would expect the effect of this process to cause an underestimate of \cVLA\ and would affect \cBal\ less since one is dealing with only the difference in the amount of scattering of the \Halpha\ and \Hbeta\  lines, rather than the absolute amount at \Halpha, as done in deriving \cVLA.
Taking the fraction of the light scattered in the forbidden lines, this position dependent effect should only be a few hundredths in the logarithm (dex).  

The second reason to expect \cdif\ to be non-zero is that when using only the optical emission (the Balmer line ratio method) one is dealing entirely with radiation that has not suffered a large amount of extinction, whereas the radio continuum to \Halpha\ method can be thrown off by the fact that some of the total radio continuum is produced by volumes behind regions that are very optically thick to \Halpha. In this case, one would be using a radio signal from all along the line of sight but \Halpha\ emission from only in front of the obscuring material and \cVLA\ would be too large, i.e. \cdif\ would be positive.  

The third reason for a difference would be variations in the ratio of total to selective extinction. Derivation of \cBal\ depends upon the reddening curve adopted, which is primarily determined by the ratio of total to selective extinction, while \cVLA\ is less dependent on this ratio.

Since the presence of \form\ in absorption indicates that near Orion-S there is a region where radio continuum emission occurs in part behind the totally obscuring cloud, we looked for significant variations in \cdif\ in the region southwest of the Trapezium where this could be done.   The results are also shown in Figure 1, where we have superimposed  the contours of large values of \cdif. One sees that there are two regions where \cdif\ is large, the northerly one nearly coinciding with the \form\ absorption and the other coinciding with the distinct obscuring feature to its south designated as the SW Cloud \citep{ode00}. This means that in both these regions optical emission lines are recording only part of the ionized gas along the lines of sight. Another way of wording this is that in these regions the model of the MIF front based on a single dominant emitting layer cannot be accurate. Since the \citet{wen95} study used the reddening corrected \Halpha\ line for reference, in these regions the model will be a depiction of the  closer (to the observer) emitting region. 
This means that the ``bump'' in the Orion-S region is there, but it is not simply the result of a nearer part of the MIF, but, it is an isolated optically thick feature directly ionized on its surface by \ori\ and behind it is another ionized region.  The more southerly region of large \cdif\ is probably similar, but it must not be as dense as the Orion-S feature since one does not see \form\ in absorption.  

\subsection{Radiation Shadows in the Orion Nebula}

Nature provides a way of probing where material is located within the Orion Nebula through the ionization shadows created beyond the proplyds.  In a study of linear features within the Orion Nebula and the Helix Nebula, \citet{ode00a} established that these features  can occur when there is a single dominant ionizing star and an intervening object optically thick to the EUV Lyman continuum (LyC) radiation. The first order theory for this situation had already been determined \citep{can98}.  Gaseous material lying beyond an object like a proplyd (in the Orion Nebula) or a knot (in the Helix Nebula) will be shadowed from photoionizing LyC photons coming directly from the ionizing star, but will be exposed to LyC photons produced through the recombination of surrounding nebular gas. This diffuse radiation field will be a small fraction of the direct radiation from the ionizing star and there will be fewer higher energy photons than in the direct radiation. The difference in the radiation field is because the recombinations occur preferentially to energy states just above hydrogen's ionization threshold.  The lower density of ionizing photons means a lower rate of photoionization and the different energy distribution means that any shadowed gas will equilibrate at a lower electron temperature.  The lower temperature means that forbidden line radiation will have a reduced emissivity and recombination lines will have an increased emissivity. This shows up particularly well when looking at the ratio of images in \oiii{} and \Halpha, as shown in Figure 2 and Figure 3. \cite{ode00a} established that  the linear rays seen in the \oiii{}]/\Halpha\ image occurred along lines of sight passing through \ori\ and a proplyd, but not all proplyds have visible radiation shadows.

\begin{figure}
\includegraphics[width=\linewidth]{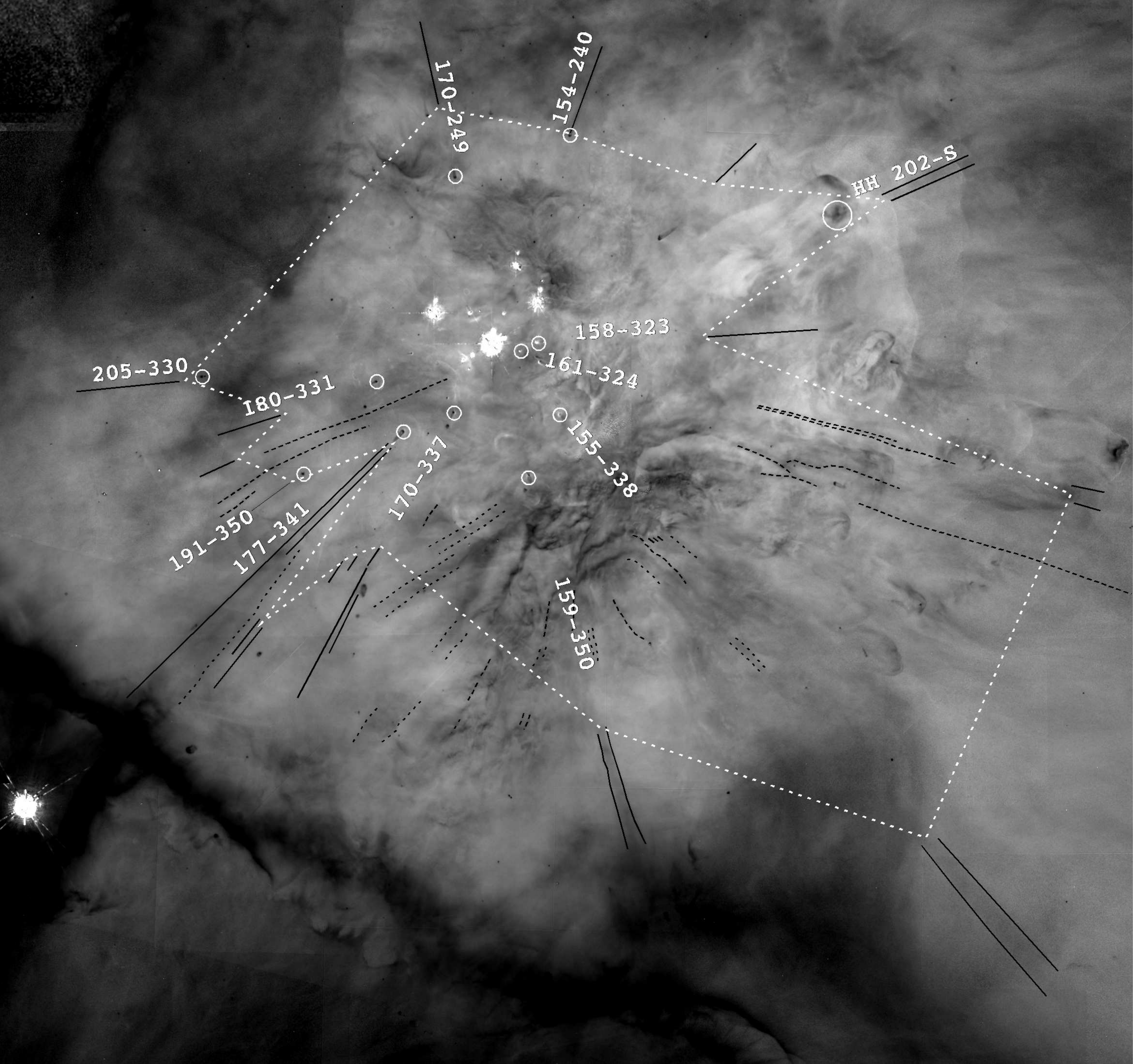}
\caption{The same field of view as Figure 1 is shown except that here the ratio of the F502N over F656N images are shown. Solid lines depict radial ionization shadow features that project back towards \ori\ and dotted lines depict significant nearly linear features that do not project back to \ori. Almost all of the solid-line features have an identified proplyd which is the occulting source and in this case the designation of the proplyd is given, adjacent to the proplyd's position. The exception to this rule is the ionization shadow feature lying beyond the HH 202-S shock. The white dashed line shows the boundary where radial shadows begin.  
\label{ratio}}
\end{figure}

These shadowed regions will only appear when there is material  present. This means that the absence of a radiation shadow indicates that only low density gas lies within that conic column. In the case where the radiation shadow does not begin at the proplyd, then the \Final{gas occupying the space between the proplyd and} the start of the radiation shadow must be of low density.  

Figure 2 has had all the linear and nearly linear features traced, solid lines being used for the indistinguishably linear features and dashed lines \Final{for} the nearly linear \Final{features} (which cannot be radiation shadows unless second order effects discussed in \citealp{ode00a} are in play).  One sees that in essentially every case of a linear feature there is an identified shadowing proplyd and, again, that not all proplyds have radiation shadows. In many cases, the associated proplyd is well inside the radiation shadow (closer to \ori).  The general pattern is clear, none of the proplyds near \ori\ have radiation shadows extending close to them and the proplyds with closely approaching radiation shadows are distant from \ori. We interpret this to mean that there is a region of low density (or very different ambient condition) gas lying within the approximate boundary shown in Figure 2. This irregular boundary in the plane of the sky is the projection along the line of sight and the central cavity is probably somewhat larger than the outlined boundary. The irregular boundary drawn indicates that there is an asymmetric  distribution about the dominant ionizing star \ori.

It should be noted that the high velocity feature HH 202-S \Final{\citep{ode08a}} \Note{Corrected reference.} also casts a radiation shadow. This indicates that this shock produces a sufficiently high density that it is optically thick to LyC photons and becomes a local ionization front within a generally ionized volume. This vitiates any argument that HH 202 was caused by high velocity material from the Orion-S region impinging on either the foreground Veil or a section of the MIF that had curved towards the observer and intercepted the collimated outflow.

\subsection{Large Scale Ionization Structures within the inner Huygens Region}

Within the paradigm of the Orion Nebula being described to first order as a relatively thin blister of ionized gas on the \ori\  side of the MIF, one expects that the \nii{} emitting zone will be quite narrow as compared with the \Halpha\  emitting zone, which includes all of the ionized gas. \oiii{} emission must arise from closer to \ori\ and over a larger volume since the effective temperature of \ori\ is insufficient to produce a He$^{++}$ zone (where O$^{++}$ has been ionized and \oiii{} is not emitted).  This means that the \nii{}/\Halpha\ ratio is a good indicator of when the MIF is highly tilted with respect to the plane of the sky. This is the reason why the Bright Bar feature is so well delineated in the left panel of Figure 3. Although there are many unexplained features of the Bright Bar, the overwhelming evidence is that it is an escarpment in the MIF where we see almost along the local ionization front.  This is probably also the correct interpretation of the low ionization zone to the southwest of the Trapezium that lies on the northeast side of Orion-S.  

\begin{figure*}
\includegraphics[width=\linewidth]{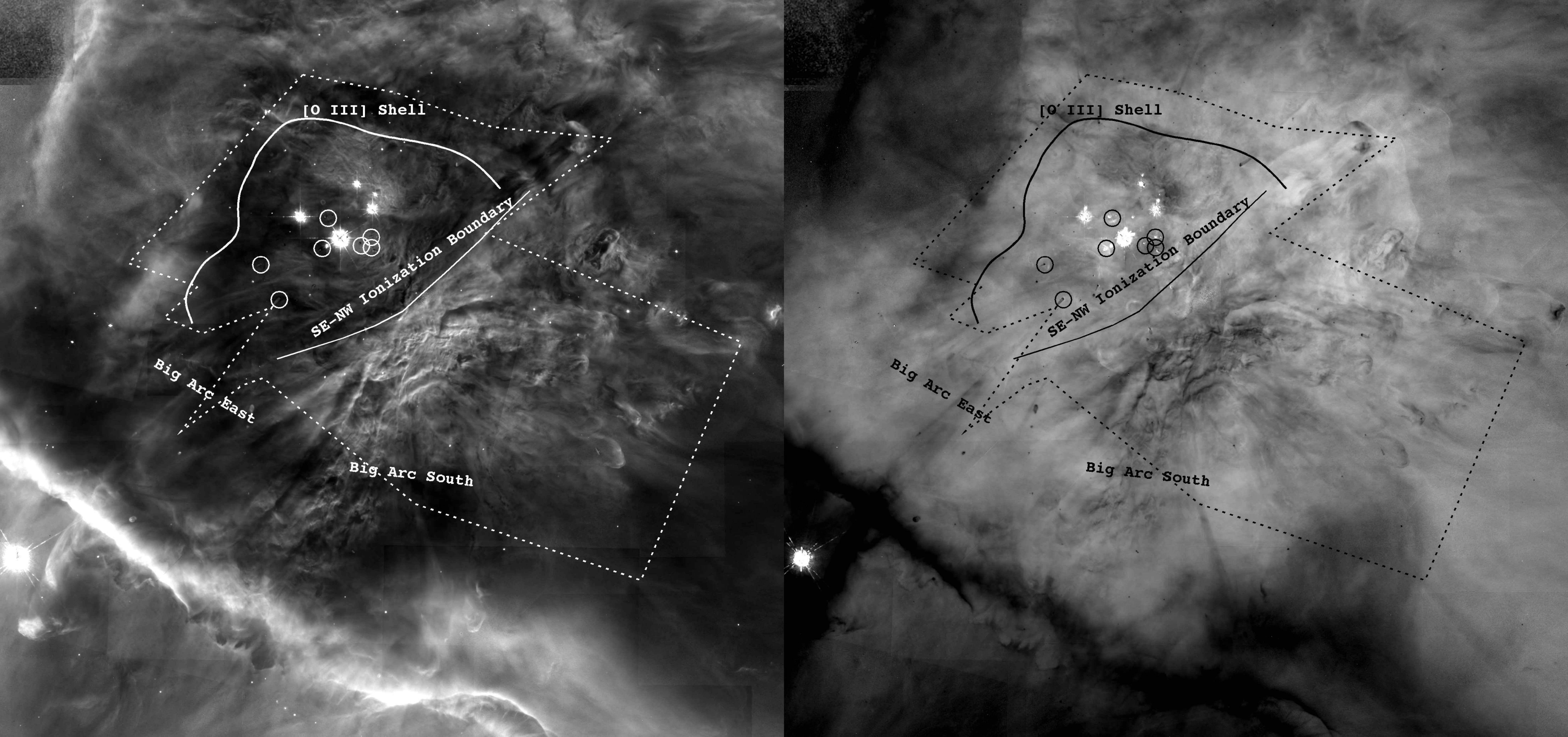}
\caption{The same field of view \Final{as} in Figures~1 and 2 is shown \Final{in both panels}. The left panel shows the ratio of F658N/F656N (\nii{}/\Halpha) and the right panel F502N/F656N (\oiii{}/\Halpha).  Both are intended to show the systematic ionization changes across the Huygens region. The left panel is little affected by local variations in interstellar reddening, while the right panel is. The dashed line indicates the boundary where radial shadows begin. The inner boundary of the \oiii{} Shell feature discussed in \S\ 2.6 is outlined with a curved solid line. The lighter solid line indicates the southwest boundary of a high ionization region close to the Trapezium stars. The circles are centered on the proplyds \Final{that have} stand-off shocks, \Final{which indicate that} they lie within the \Final{unshocked hypersonic stellar wind} from \ori.
\label{twopanels}}
\end{figure*}

\subsubsection{A Previously Unrecognized High Ionization Structure within the Inner Huygens Region}

Figure 3 \Final{(left panel)} reveals an unexpected large scale structure in a C-shaped zone of low \nii{}/\Halpha\ with \ori\ located near its middle.  Close examination of Figure 3 \Final{(right panel)} indicates that where \nii{} emission is weak, \oiii{} emission is strong. This means that the C shaped zone is an ionization phenomenon.  The fact that the \nii{} deficit and \oiii{} excess is an irregular line, rather than a closed area, indicates that this is a shell of high ionization. Henceforth we will call this the \oiii{} Shell, with full recognition that this feature is probably not closed and we do not understand its three dimensional characteristics. South of the Trapezium and Orion-S lies a large high ionization feature labeled as Big Arc South and Big Arc East. This was originally discovered in a low velocity resolution study \citep{ode97}, then included in a later higher resolution study \citep{doi04}, and the correct interpretation as a partial shell blueshifted about \Vsun\ = 9 \kms\ being presented in \citet{gar08} and confirmed by our remeasuring a sample of the Big Arc from the \oiii{} spectra data set \citep{doi04}. \Will{Although
  the Big Arc is seen in projection to be almost continuous with the
  southern arm of the \oiii{} shell, it is very distinct
  kinematically, and the physical relation between these two features
  remains unclear.} 

The approximate alignment of the boundaries of the radiation shadows
and the \oiii{} Shell argues that this is an incomplete shell with
lower central density within the larger mass of ionized gas. It is not
an ionization boundary because one does not see \nii{} enhancement on
its outer perimeter. The feature seems to open to the west-southwest,
being most open in the direction of Orion-S. There is a nearly linear
border to the high ionization zone, which \Will{we have} labeled in
Figure 3 as the SE-NW Ionization Boundary. It is not immediately
obvious that this feature is an ionization boundary or caused by the
higher extinction of the foreground Orion-S Cloud (cf.~\S~3.1)

Figure 3 also shows the boundary of the radiation shadows from Figure 2 and we see that there is good agreement of this boundary with the \oiii{} Shell, which further strengthens the argument that this \oiii{} excess traces the outer boundary of a volume of low gas density. We also show in Figure 3 the location of proplyds with stand-off shocks and find that all of them fall within the C-shaped \oiii{} Shell. Since it is most likely that these stand-off shocks are formed by direct exposure to the high velocity stellar wind of \ori, this argues that the \oiii{} Shell is created by that stellar wind, an idea developed in \S\ 3.1.

\subsubsection{Physical Conditions in the [\ition{O}{3}] shell}
\label{sec:phys-cond-oiii}
The average hydrogen number density of the \oiii{} shell, \(n\), can
be estimated from the H\(\alpha\) surface-brightness
(\sbunits), given by
\begin{equation}
  \label{eq:sha}
  S(\mathrm{H\alpha}) =  x\elec \alpha\eff n^2 L / 4\pi ,  
\end{equation}
where \(L\) is the path length of the line of sight through the shell
and \(x\elec\) is the electron fraction. Both hydrogen and helium are
assumed to be singly ionized, with \(\mathrm{He/H} = 0.0977\)
\citep{1998MNRAS.295..401E}, so that \(x\elec = 1.0977\). The
effective line recombination coefficient \(\alpha\eff\) has a very
weak density dependence and can be approximated in the Case~B limit as
\(1.16 \times 10^{-13} T_4^{-1} \mathrm{\ cm^3\ s^{-1}}\)
\citep{2006agna.book.....O}, where \(T_4 = T / 10^4\mathrm{\ K}\). For
a spherical shell of radius \(R\) and fractional thickness \(\Delta
\ll 1\), the maximum path length through the shell is \(L \simeq 2
\Delta^{1/2} R\). The observed radius and thickness of the \oiii{}
shell vary with PA by about 50\% but we will take \(R = 2 \times
10^{17}~\mathrm{cm}\), \(\Delta = 0.28\) as typical of the
well-defined northern arm of the shell, so that \(L \simeq R\).

The surface brightness of the shell is most easily estimated from the
sharp brightness jump at the inner edge of its limb-brightened
northern arm. Using WFPC2 images
\citep{ode96}, flux-calibrated following
\citet{ode99} and extinction-corrected following
\citet{ode00}, we find \(S(\mathrm{H\alpha}) = 2.85
\times 10^{10}~\sbunits\), which represents about 25\% of the total
nebular surface brightness at the shell position. We also find
\(\oiii{}/\mathrm{H}\alpha = 1.62\) and no evidence of any \nii{}
emission from the shell (\(\nii/\mathrm{H}\alpha < 0.01\)).

For the shell temperature, we use the results of
\citet{2008ApJ...675..389M} who find \(T \simeq 8300~\mathrm{K}\) at
the position of the shell from the ratio of the 4363~\AA{} and
5007~\AA{} \oiii{} lines (upper-right panel of their Fig.~6). This
temperature represents an average of the emission from the shell and
the background nebula at that position, but there is no evidence for
any significant jump in \(T\) at the inner edge of the shell, so we
can be confident in using it for the shell temperature, giving an
effective recombination coefficient of \(1.36 \times 10^{-13}
\mathrm{cm^3~s^{-1}}\).


Combining all the above, we find a shell emission measure of
\(\mathrm{EM} = x\elec n^2 L = 2.56 \times 10^{24}~\mathrm{cm^{-5}}\)
and hence a density \( n = 3400~\pcc\). The ionizing flux at the
position of the shell is \(\Phi_{\mathrm{H}} = Q_{\mathrm{H}} / 4 \pi
R^2 \simeq 3.6 \times 10^{13}~\fluxunits\), where \(Q_{\mathrm{H}}\)
is the ionizing luminosity of \thC{}, assumed to be \(1.8 \times
10^{49}~\mathrm{s}^{-1}\) \citep{2005ApJ...627..813H}. 

The ionization parameter in the shell (\(\Phi_{\mathrm{H}}/ n c =
0.35\)) is very large when compared with typical values for the nebula
as a whole (0.01--0.02). Therefore, if dust grains are not
significantly underabundant, the grains will dominate the ultraviolet
opacity of the shell. Assuming a dust extinction cross-section of
\(\sigma_\mathrm{d} = 10^{-21}\ \mathrm{cm^2}\) per H nucleon
\citep{bal91}, the optical depth of grains in the shell
to ionizing radiation is \(n \sigma_\mathrm{d} R \Delta \simeq
0.2\). For comparison, the neutral hydrogen optical depth is much
smaller: \(\alpha_{\mathrm{B}} \mathrm{EM} / \Phi_{\mathrm{H}} \simeq
0.022\). Thus, the shell is optically thin to EUV radiation, which, combined with the high ionization parameter, means that low-ionization lines such as [\ion{N}{2}] are expected to be very weak. Indeed, no detectable [\ion{N}{2}] emission is observed from the shell.


\subsection{Space Motion of \ori}

In spite of its brightness, it has been difficult to determine the \Final{three-dimensional} space motion of \ori{}, the dominant ionizing star. This is true for both the tangential and radial velocity components that are needed for deriving \Final{this} motion.

The brightness relative to the other cluster members presents particular challenges in determining the tangential velocity. The most accurate study is that of \citet{van88}, who determined that \ori\ was moving at \(4.8\pm 0.5~\kms\) towards Position Angle 142\arcdeg\ and that this was significantly larger than the dispersion value of \(1.5\pm 0.1~\kms\) found for the other cluster stars (49 cluster members in the sample). \citet{jon88} did not include the brightest members of the ONC in their astrometric study, but did find a \Final{velocity} dispersion of \(1.5 \pm 0.7~\kms\) for the brighter stars in their study.  \citet{tan04} interpreted this high tangential velocity away from the direction of the imbedded BN-KL cluster stars as evidence that \ori\ shares a point of origin with the radio source BN, which is moving \Final{in} the opposite direction at a high velocity (\(38~\kms\) at \(\mathrm{PA}=322\arcdeg\)), and that the present tangential velocity would have placed the BN object in the vicinity of the Trapezium about 4000~years ago. However, subsequent studies \citep{rod05,gom05,gom08} of the radio sources within the BN-KL region show that three radio sources in the center of the BN-KL region are moving away from a common center with a timescale of 500 years and that the BN-KL source I used for reference by \citet{tan04} was one of the moving objects.  \citet{gom05} found that BN was moving towards the northwest at 24 \kms\ in the Orion cluster rest frame. The separation of \ori\ and the dynamic center of BN-KL is 63\arcsec, which means that a backwards projection of \ori 's tangential motion as determined by \citet{van88} would cross that position in about 28,000 years. Even if it is moving away from the BN-KL center, this timescale indicates that it is unlikely \ori\ is a runaway object from the BN-KL region event that produced the three escaping radio sources and the bipolar outflow with an upper limit age of 1000 years \citep{doi02} that has produced the well known \htwo\ fingers.

Even the modest anomalous tangential velocity of 4.8 \kms\ may be too high.  The data of \citet{van88} indicate that \ori\ and $\theta ^{1}$Ori~B are moving in opposite directions at about 3.5 mas per year. In a compilation of visual astrometric measurements of the relative positions of the Trapezium stars, \cite{all74} present results for these two stars over a time interval of 125 years, during which time the stars would have separated by 0.43\arcsec , but, there is no detectable change to an accuracy of about 0.15\arcsec. 

Determination of the radial velocity of \ori\ is also difficult \citep{ode01b,vit02}. In this case the brightness of the star aids the collection of high velocity resolution spectra, but the star is complex, being an oblique-rotation magnetic star which shows a bewildering pattern of periodic and erratic spectral changes. It has a resolved early type star companion. Properities of the system are summarized in \cite{mue08}.  Utilizing only spectroscopic radial velocity data, \citet{vit02} determined a systemic radial velocity of \(10.9 \pm 2~\kms\) and by combining spectroscopic and astrometric \citep{kra07,pat08} data, \citet{sta08} have determined the systemic radial velocity of \ori\ as \(13~\kms\), which is quite different from that of the other Trapezium stars and the lower mass ONC members, as summarized in Table 1, and indicating that it is moving rapidly away from the OMC and its host star cluster (relative velocity about \(13~\kms\)).
\Final{In \S~\ref{sec:what-impact-putative} below, we examine the effect that such a rapid motion for \thC{} would have on} the present structure and evolution of the Orion Nebula.

This special characteristic (high spatial velocity) may not be all that unexpected. \citet{sto91} concludes that in general 40\% of all O stars are runaways, although only one in ten OB stars has a known binary companion.  There is an observational selection effect operating in the ONC that favors detection of blueshifted radial velocities. A redshifted runaway would be moving into the obscured portion of the OMC and we would not be seeing the photoionized Orion Nebula \Will{at visual wavelengths}. This means that if we are going to detect an anomalous radial velocity, it is going to be a blueshift. The difference in radial velocity with respect to the other Trapezium stars would mean that either this juxtaposition of massive stars is coincidental or perhaps that dynamical interactions with the other \Final{stars} has produced \ori{}'s anomalous radial velocity. 

Certainly, runaway stars are expected from a grouping like the Trapezium \citep{pak06} and have been created previously in the ONC\@. The divergence from Orion of the two high velocities stars AE Aur and $\mu$ Col has been known for over a half century \citep{bla54}. Calculations of the trajectories of these stars and $\iota$ Ori \citep{hoo01} indicate that they all originated about 2.5 million years ago from a position now occupied by the ONC. This is not to argue that these stars arose in the Trapezium grouping, however, it is evidence that there was an earlier epoch of massive star formation in this vicinity. Given today's presence of three centers of star formation (the ONC, BN-KL, and Orion-S), it is not hard to accept that there was at least one earlier similar center. Likewise, the earlier creation of runaways argues that this could have again occurred and \ori\ is the product.



\section{Discussion}

\Bob{The new features presented in \S\ 2 have caused us to reassess several major properties of the Orion Nebula. In \S\ 3.1 we consider the effects of the high velocity wind from \ori. In \S\ 3.2 we present arguments for the Orion-S feature being a separate cloud within the ionized cavity of the nebula and in \S\ 3.3 present a model consistent with its producing \form\ in absorption. We conclude with a discussion of how our view of the nebula would be altered if \ori\ has a large blueshifted velocity with respect to the other members of the ONC.}

\subsection{The Effects of the High-Velocity Wind from \ori}

The presence of a high velocity stellar wind from \ori\ is characteristic of early spectral type stars and will certainly affect the conditions of the nearby nebular gas. In this section we address the nature of the wind, its expected interactions, and compare the expectations with what is observed.

\Will{The} radiatively driven stellar wind is modified by the star's
strong dipolar magnetic field \citep{2002MNRAS.333...55D}, leading to
a very anisotropic outflow \citep{2005PASP..117...13S}. However, the
large inclination between the magnetic and rotation axes
\citep{2006A&A...451..195W} means that much of this anisotropy is
smoothed out over the 15.422 day rotational period
\citep{1993A&A...274L..29S, 1996A&A...312..539S}, so that it is
reasonable to assume that the wind is approximately isotropic when
considering its interaction with the Orion Nebula. Angle-averaged
values for the mass-loss rate and terminal velocity of \(\dot M\wind =
4 \times 10^{-7}~\smyr\) and \(V\wind = 1400~\kms\) have been
determined from modeling the X-ray emission from the base of the wind
\citep{2005ApJ...628..986G}.

A totally independent measurement of the strength of the fast stellar
wind is provided by the \oiii/MIR arcs seen in front of the closest
proplyds to \thC{} \citep{1994ApJ...433..157H,
  bal98,rob05, smi05}, which
are interpreted as the stationary stand-off bowshocks that result from
the interaction of the proplyd photoevaporation flow with the stellar
wind \citep{2001ApJ...561..830G}. The position of these bowshocks is
determined by the balance between the ram pressures of the proplyd
outflow and the fast wind from \thC, so that they can be used as
probes of \( \dot M\wind V\wind \) so long as the proplyd parameters
can be determined with sufficient precision. Application to the
best-studied proplyd 167-317 \citep{2001ApJ...561..830G,
  2002RMxAA..38...51G, hen02} yields a value of \( (\dot
M\wind / 10^{-7}~\smyr) (V\wind / 1000~\kms) = 4.6 \pm 1.4\), which is
consistent with the \citet{2005ApJ...628..986G} values given above.

\subsubsection{\Final{Theoretical Predictions of the Wind/Nebula Interaction}}

\begin{figure*}
\includegraphics[width=\linewidth]{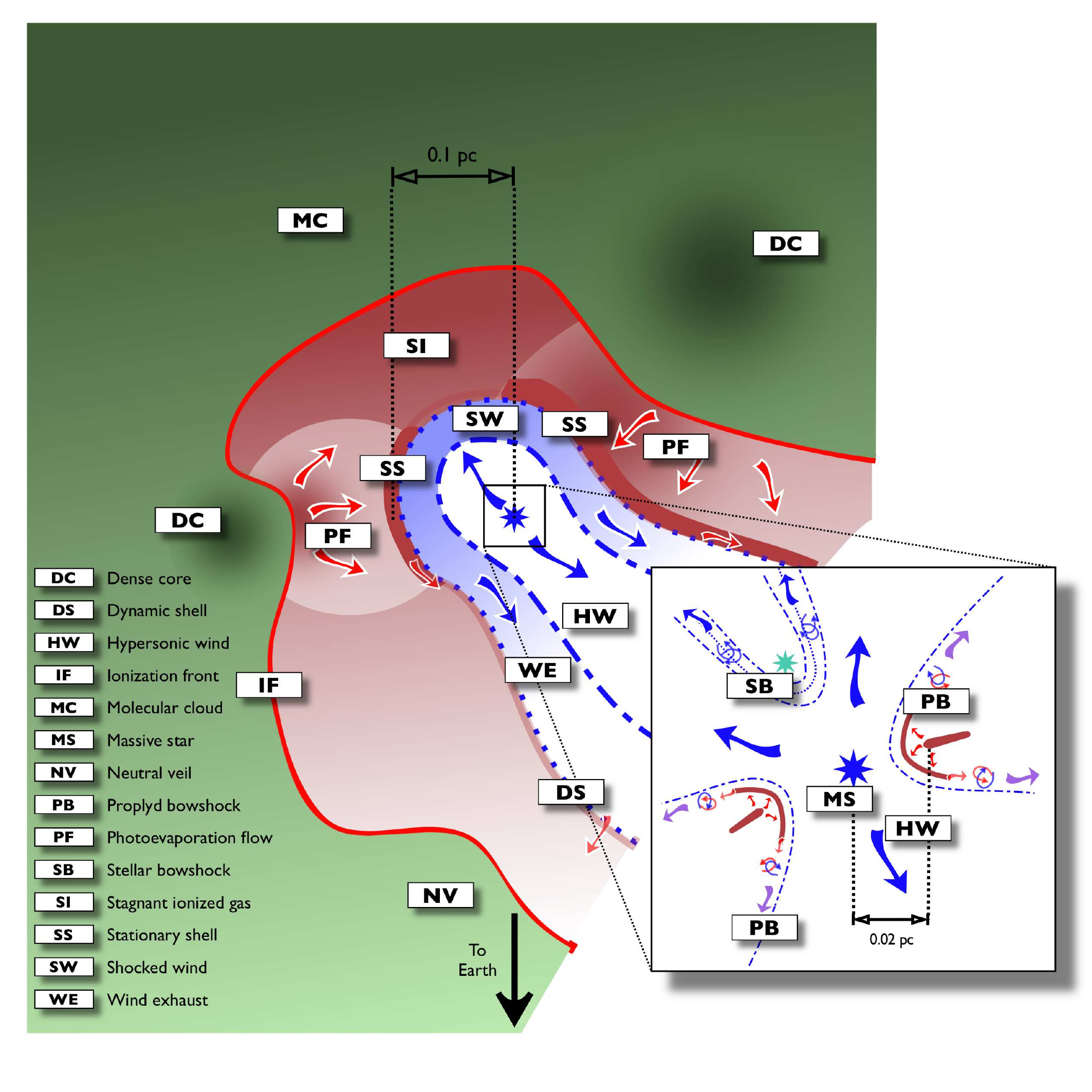}
\caption{Schematic cross-section through the inner Orion Nebula,    showing the different zones of stellar wind interaction. Blue indicates stellar wind material, red indicates photoionized    nebular material, and green indicates atomic/molecular gas, with darker shading corresponding to qualitatively higher density within each region. Arrows show the approximate direction of gas flows, with sizes that roughly scale with the momentum flux. Shocks in the stellar wind are shown by short-long-dashed lines, while the contact discontinuities are shown by dotted lines.  The drawing does not represent any particular cut through the nebula, and is not exactly to scale, although the sizes in  parsecs of some major features are marked. The observer is located off the bottom of the picture. The inset box shows a zoom of the central region.
\label{WillsFigure}}
\end{figure*}

The interaction of the fast wind from \thC{} with its surroundings differs from the standard spherical hot-shocked wind bubble scenario \citep[see e.g.,][]{1997pism.book.....D} due to the effects of geometry, density gradients, mass-loading from embedded sources, and possibly thermal conduction. The different wind interaction regions are summarized in Figure~4, which synthesizes findings from many recent theoretical and observational studies. Each region is discussed further below, starting from the inside. We concentrate on the inner parts of the nebular that are relevant to the observational material presented in this paper, leaving aside considerations of the EON from which x-ray evidence of stellar wind interaction has recently been reported \citep{gud08}.

\paragraph{Mass loading zone}
Mass loading of a stellar wind by embedded sources lowers the stellar wind velocity of the unshocked wind and consequently reduces the temperature and modifies the cooling properties in the hot shocked wind region \citep{1986MNRAS.221..715H}. In the Orion Nebula, the sources of mass are the photoevaporated flows from the close-in proplyds (PB in the figure) and the stellar winds from the other Trapezium stars (SB). Of these sources, the proplyds are the most important \citep{2001ApJ...561..830G}, since the mass-loss rate from each proplyd  is of the same order as \(\dot M\wind\) \citep{1999AJ....118.2350H,hen02}, whereas the most massive of the other Trapezium stars are early B-type, with winds at least an order of magnitude weaker than that of \thC{} (MS in the figure). Unfortunately, the efficiency of the mixing between the stellar wind and the proplyd photoevaporation flows is poorly constrained \citep{2002RMxAA..38...51G}, leading to large uncertaities in the degree of deceleration of the wind. However, as long as the mixed wind remains highly supersonic, its momentum \(\dot M\wind V\wind\) is conserved.

\paragraph{Shocked wind}
When this mass-loaded, but still supersonic, wind interacts with the
nebula, a two-shock flow pattern is formed \citep{1968ApL.....2...97P,
  1972A&A....20..223D}: an outer shock that accelerates the nebular gas and an inner shock (blue double-dashed line in Figure 4) that decelerates the stellar wind. The two shocked regions are separated by a contact discontinuity (blue dotted line), across which the pressure is continuous. If the inner wind shock is non-radiative (as is usually the case), the shocked stellar wind (SW) will have a temperature \(T / 10^6~\mathrm{K} \simeq 0.15 (V\wind/100~\kms)^2\) and a thermal pressure roughly equal to the ram pressure at the inner shock.

Large-scale density gradients in the adjoining molecular cloud (MC)
mean that the inner shock is much closer to the star in the cloud
direction (upper-left in Figure 4) than in the
opposing direction (off-picture to lower-right), leading to a higher
pressure of the shocked stellar wind in the cloud-facing direction,
which therefore flows towards the low-density end of the bubble
\citep{2006ApJS..165..283A} in a transonic wind exhaust (WE in
Figure 4). A simple steady-state model then gives
the radius of the wind shock as 0.73 times the radius of the
contact discontinuity.

\paragraph{Shocked nebula}
In the early evolution of an \hii{} region, the ionization front is
likely to be trapped inside the dense shell swept up by the outer
wind-driven shock \citep{1996ApJ...469..171G, 2006ApJS..165..283A}.
The density in the shell decreases with time, so that the ionization
front eventally escapes from the shell and drives a low-velocity
neutral shock ahead of itself into the molecular gas
\citep{2007dmsf.book..183A}. After this time, the outer wind-driven
shock will be interacting with photoionized nebular gas, as shown in
Figure~4.  If this gas is static, with a constant
hydrogen number density \(n~\pcc\), then the outer shock will
degenerate into a sound wave after \(\simeq 1.3/n^{1/2}\) million
years \citep{1997pism.book.....D}, leading to dissipation of the
shocked nebular shell, after which one will have approximate static
pressure equilibrium between the \hii{} region and the shocked
wind. In the cloud-facing direction, where the ionized density is \(n
= 10^3\)--\(10^4~\pcc\), this will occur on a timescale much shorter
than the age of the nebula, so that stagnant zones of ionized gas (SI
in Figure 4) can be confined by the wind
\citep{2006ApJS..165..283A}. In the direction away from the cloud, the
ionized density is much lower, \(n < 100~\pcc\)
\citep{1993A&AS...98..137F}, so that one may still have a dynamic
shell (DS) of swept-up \hii{} region that moves outward at a few tens
of \kms.

Additional structures will arise due to the internal dynamics of the
ionized gas in the \hii{} region. Turbulence and gravitational
collapse in molecular clouds \citep{2005ApJ...618..344V} produce
density concentrations on multiple scales (idealized as dense cores,
DC, in Figure 4), which retard the progress of the
ionization front. In regions where the ionization front is locally
convex or flat, photoevaporation flows (PF) will develop
\citep{2005ApJ...627..813H, 2006ApJ...647..397M}, which reach mildly
supersonic velocities of 20--40~\kms{} \citep{1968Ap&SS...1..388D,
  1989ApJ...346..735B}. Where one of these photoevaporation flows
interacts with the stellar wind bubble, a stationary shell (SS) of
shocked ionized gas will form at a position where the ram pressure of
the photoevaporation flow balances the thermal pressure of the shocked
wind.

\subsubsection{Observational Evidence for the Effects of the Wind on
  the Ionized Nebula}

As reported in \S~2.6.1 above, we have found an incomplete high-ionization shell in the inner Orion Nebula, which may be related to the action of the stellar wind from \thC{}. Since the shell has both an inner and outer boundary (Figure 3), it would correspond to a stationary shell in the context of the above discussion (SS in Figure 4). This means that the kinematics of the shell is likely to be unremarkable, as compared with the kinematics of the general \oiii{} emission from the nebula. Just as with the rest of the nebular gas, the gas in the shell will find it easier to move in the direction away from the molecular cloud and so should be slightly blue-shifted on average compared with the systemic velocity of the stars and molecular gas. Inspection of the \oiii{} velocity cubes of \citet{doi04} shows that the line profiles from the shell are, indeed, very typical of the nebula as a whole.

\newcommand\sh{_{\mathrm{sh}}} \newcommand\bub{_{\mathrm{bub}}} 

The inner boundary of the shell corresponds to the contact discontinuity with the shocked wind bubble, so the shell thermal pressure should equal that of the shocked wind. From the physical parameters derived from observations in \S~\ref{sec:phys-cond-oiii}, one finds a shell pressure of \(P\sh/k = (1 + x\elec) n T \simeq 5.9 \times 10^7~\pcc\ \mathrm{K}\). The bubble pressure will be equal to the wind ram pressure at the inner wind shock: \(P\bub/k = \dot M\wind V\wind / 4 \pi R_{\mathrm{in}}^2 k \simeq 7.84 \times 10^7~\pcc\ \mathrm{K}\), where we have assumed that the inner shock radius is \(R_{\mathrm{in}} = 0.73 R\) (see above) and that the wind momentum loss rate \(\dot M\wind V\wind \) is the proplyd-derived value discussed in \S~3.1. Given the uncertainties in their derivation, these pressures are indistinguishably the same, providing strong evidence that the \oiii{} shell has been compressed by the stellar wind.

The Red Bay \citep{gar07} is a region in which the high-ionization gas, as traced by [\ion{S}{3}] and \oiii{} lines, has a velocity similar to that of the molecular cloud (\(V_\odot = 25\)--\(30~\kms\)), rather than the more blueshifted velocities (\(V_\odot < 25~\kms\)) found elsewhere in the nebula. The region of relatively redshifted velocities starts at the Trapezium and extends in a broad strip (\(\simeq 75'' \times 25''\) ) to the east-southeast, which can be best appreciated as a light-colored region in the \oiii{} mean velocity map (Fig.~13 of \citealt{gar08}).\footnote{As originally
  drawn in Fig.~7 of \citet{gar07}, based on relatively low
  signal-to-noise [\ion{S}{3}] observations, the Red Bay was broader,
  extending somewhat north of the Trapezium. However, this
  interpretation is not supported by the better-quality \oiii{} data.}
The Red Bay is also associated with a region of lower-than-average
\oiii{} surface brightness.

Comparison of the emission measure and electron density in this region \citep{gar07} shows that it is \textit{not} a thin shell seen face-on, but has a thickness along the line of sight \(\simeq 10^{18}~\mathrm{cm}\), which is roughly twice its extent in the plane of the sky. We therefore suggest that the Red Bay may correspond to a stagnant zone of ionized gas (SI in Figure 4), which is trapped between a concave ionization front, the stellar wind bubble, and the photoevaporation flows from molecular concentrations such as the Bright Bar.

\subsection{The 3-D Structure of the Orion-S Region}

The several pieces of published and newly derived evidence require that we revise the accepted picture that Orion-S is simply a local rise in the MIF caused by dense underlying material in the OMC\@. Certainly, this is a region of enhanced density as determined from H$\rm ^{13}$CO$^{+}$ observations. The recently published 21\arcsec\ resolution H$\rm ^{13}$CO$^{+}$ maps \Final{that include} the Huygens region \citep{ike07} show the most massive cores to lie about 30\arcsec\ north of BN-KL and within Orion-S, but these observations cannot discriminate between cores that lie within a foreground optically thick feature and something that lies behind the MIF and within the \Will{main OMC filament}. However, the presence of \form\ absorption in the radio continuum (discussed in \S\ 2.3) clearly places a high density, very optically thick object in front of an ionized volume and this \FFinal{object could also be the source of the peak in rotationally excited CO emission that lies between the \form{} absorption feature and the Trapezium \citep{wil01}.}
The disparity in the extinction derived from radio to optical and optical line ratios (discussed in \S~2.4) further confirms that the optical features we see lie in the foreground of an ionized volume. The low ionization seen on the northeast side of Orion-S is most easily explained as an inclined face analogous to the escarpment causing the Bright Bar, but not as tilted. Otherwise it is impossible to explain the \nii{} surface brightness this close to the ionizing star.

There are arguments that the foreground Orion-S Cloud hosts the embedded stars that are the sources of several optical features and outflows. A dark rectangular form lies almost exactly on the redshifted SiO outflow and the infrared source 139-357. It is not certain which (or if both) of these two features is associated, but an association of an optical feature with either places that feature in the foreground, rather than something within the OMC.   There is a suggestive notch in the \form\ absorption contour coincident with both the blueshifted outflows in SiO and CO, so that one or both of these must be in the foreground feature. The presence of high ionization, high tangential velocity features symmetric with the Dark Arc \citep{ode00,ode08b} indicate that it is also part of the foreground structure.

The \form\ absorption indicates very high optical depths in the near ultraviolet and it is unlikely that the obscuration by the foreground feature stops at the lowest contour of detection of \form\ absorption.  This high extinction allows us to comment on the location of the sources that give rise to optical outflows. The sources of the large scale optical outflows have been isolated to a small region shown as the dark dashed-line ellipse in Figure 1 \citep{ode08b}. These outflows are all blueshifted. The east-oriented HH~529 shocks are high ionization and are seen very close to the source, indicating that this flow emerges quickly from behind the local ionization front. The west-oriented HH~269 shocks and \Final{jet} also emerge close to their sources and are seen visually, arguing that they \Final{too} originate in the foreground Orion-S Cloud.  The emergence further from the source region \citep{ode08b} of the high velocity gas driving the northwest-oriented HH~202 and the southeast-oriented HH~203 and HH~204 could either indicate that their sources are more deeply imbedded in the Orion-S Cloud or lie within the background OMC. The latter interpretation would \Will{imply} that the lack of visibility near the source region is due to the high optical extinction of the Orion-S Cloud, but it seems more likely that \Final{all the sources lie at a similar distance along the line of sight}.

There is a question about the location of the strongest infrared and radio sources in this region: 137-408 (CS~3 in earlier, low resolution studies) and 134-411 (FIR~4 in earlier, low resolution studies). The symmetry axis of the nearby HH~530 shocks points to either of these sources and the proximity of such an optical feature argues that its source lies in the foreground \Final{Orion-S cloud}.  

The far-infrared emission from dust heated by embedded strong infrared sources \citep{hou04} indicates (Figure 1) a peak in the vicinity of BN-KL and a weaker elongated peak in the Orion-S feature. The northerly extension of this southern feature coincides with the source region for the optical outflows and the southerly extension coincides with the 137-408 and 134-411 sources. Unfortunately, this doesn't tell us if the southerly extension belongs to the foreground feature or lies within the OMC.

\begin{figure}
\includegraphics[width=\linewidth]{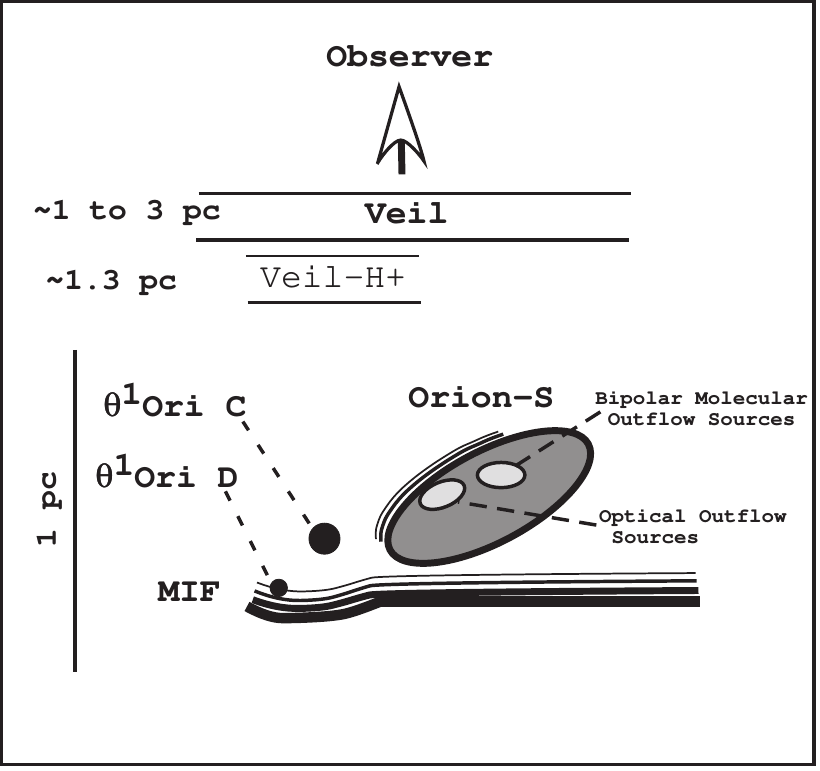}
\caption{This sketch shows a southwest cross section passing through the Trapezium and Orion-S. The MIF is depicted as flat behind Orion-S only because of a lack of information as to its true shape there. Although dimensions are shown, the uncertainty in the exact positions means that the figure is not drawn to scale.
\label{CrossSection}}
\end{figure}

The most likely geometry for this region is shown in Figure~\ref{CrossSection}. In this model we have placed the Orion-S Cloud far enough into the foreground that the MIF extends continuously behind it. The exact placement of the Orion-S Cloud in the foreground is uncertain, but it must be close to the distance of \ori\ in front of the local MIF because if it were much further from the observer it could not be distinguished from the MIF.  \Bob{In the same way that the Bright Bar probably represents where the ionization front of the nebula encounters a long high-density condensation within the host OMC, it is likely that the Orion-S Cloud was a high density feature within the OMC, which has been isolated as the ionization front progressed into the host molecular cloud. This conclusion is compatible with the fact that the velocity of the \form\ extinction feature (\Vform =28.6$\pm 2$) is essentially the same as the host molecular cloud (\Vomc =25.9$\pm 1.7$).}

\subsection{Chemistry of the Orion-S Cloud}
\label{sec:chemistry-orion-s}

Distinctly different heating processes are important in various regions of the Orion environment. Starlight is the most important heating/ionization process in the bright H$^{+}$ region and its associated PDR. Wind-driven shocks are responsible for much of the geometry and emission of the BN object. Cosmic rays and heating by dissipative MHD waves may be important in deeper regions of OMC1. Given the assortment of possible dynamical and microphysical processes involved, two approaches to numerical simulations of the ISM can be taken.  Hydrodynamics codes follow the motions of the gas but must compromise the microphysics to solve the problem on today's computers. A microphysics code (like Cloudy, which we use here) will not compromise on the microphysics but must do something simple for the structure of the cloud. Atomic processes are usually much faster than dynamical timescales so the physics that determines the spectrum is often not affected by the dynamics and a static geometry can be assumed.

In this section, we wish to understand several overall features of the Orion-S cloud via \Gary{a simulation of the ionization, thermal, and chemical state of the cloud}. The primary question we want to answer is ``Can we explain the presence of H$_{2}$CO in absorption?''. Furthermore, can we do so in the context of other molecules seen in emission and the geometrical constraints imposed by the projected size of Orion-S?  There are estimates of the gas column density, particle density, temperature, and gas pressure (which is surprisingly high). Can starlight from the Trapezium cluster account for these general properties, or are other processes at play?  Here we test whether starlight from the Trapezium cluster can account for several of the observed properties of Orion-S\@.

\subsubsection{Observational Constraints {\&} Geometry}
\label{sec:observ-constr}
The most important constraints to modeling the Orion-S cloud are the H$_{2}$CO absorption measurements, the total column density, and the projected size of Orion-S.  In cold, dark, molecular clouds, chemical models which include the effects of accretion of molecules onto grain surfaces \citep{has92,has93} show that H$_{2}$CO will reside almost exclusively on the surface of grains (henceforth ``freeze out'') at temperatures \Gary{at or below} 10~K\@.  A temperature of 10~K is also consistent with the ionized and molecular gas in Orion-S being in gas pressure equilibrium (see below).  However, since H$_{2}$CO is observed in the gas phase in Orion-S, the temperature must be higher, such that the rate of evaporation of H$_{2}$CO \Will{from} grain surfaces, \(k_\mathrm{evap} = 1.7\times10^{12} \exp(-1760/T_\mathrm{d})\), must be large enough \Will{to produce a significant gas-phase abundance of H$_{2}$CO.}
\begin{figure}
  \centering
  \includegraphics[width=\linewidth]{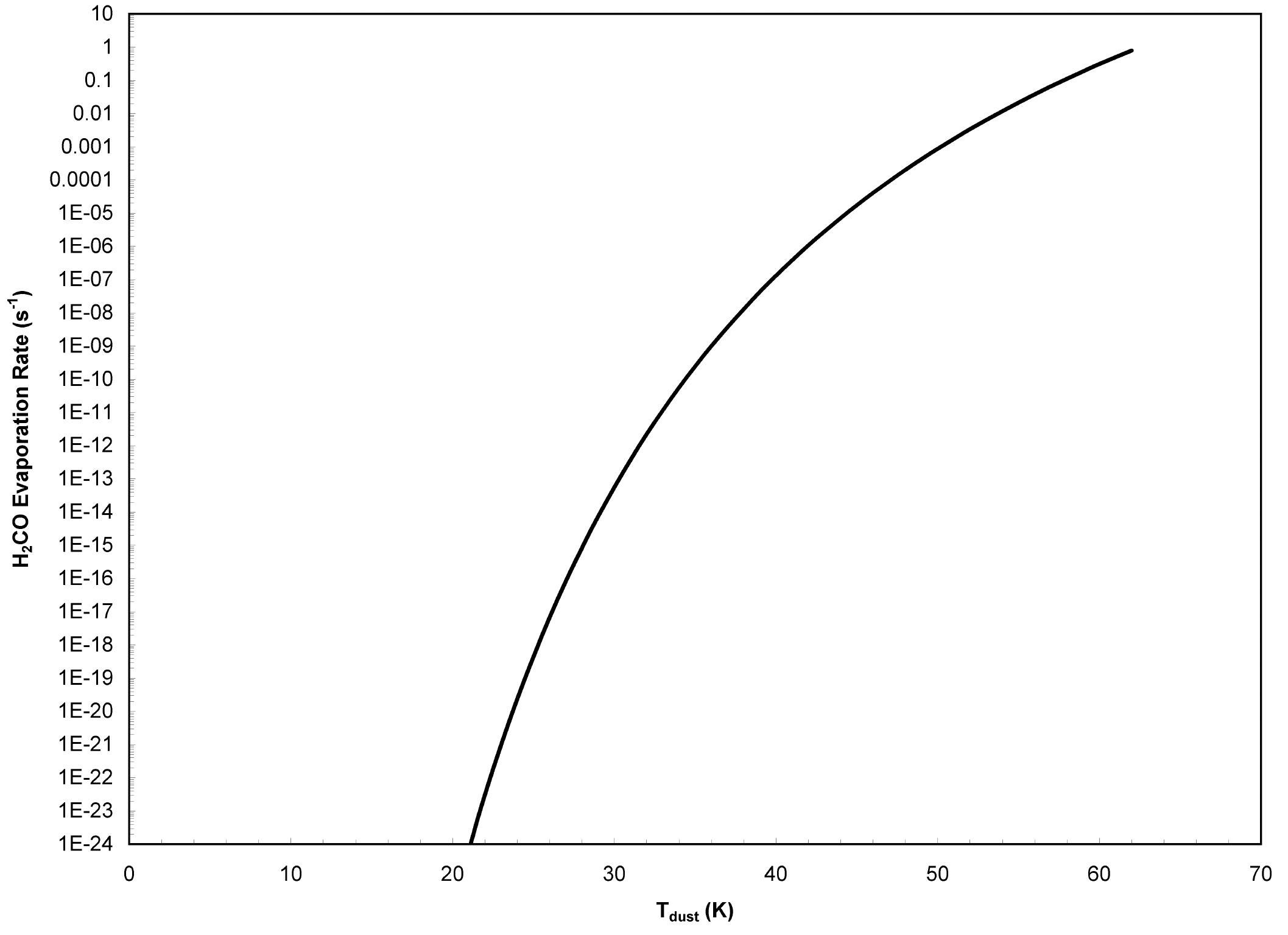}
  \caption{H\(_2\)CO evaporation rate as a function of grain temperature.}
  \label{fig:Orion-S-H2CO}
\end{figure}
Figure~\ref{fig:Orion-S-H2CO} shows that this rate will be fast (\(> 10^{-10}~\mathrm{s}^{-1}\)) for \(T_\mathrm{d} > 34 \mathrm{\ K}\).  Therefore, H$_{2}$CO observations imply that the dust temperature (\Gary{which should be nearly in equilibrium with the gas temperature for deeper regions of the cloud}) needs to be roughly 30~K or higher in order to explain the presence of H$_{2}$CO.  \Nick{The total H column density is given in \citet{mcm93}, who through 1.3 and 3.1 mm dust continuum emission found \(N(\mathrm{H}) \sim 4\times 10^{24} \mathrm{\ cm^{-2}}\) for the region 12\arcsec{} N of the 3.1~mm continuum peak, corresponding to a region \Final{they} referred to as the local quiescient cloud. The projected separation of the Trapezium stars from Orion-S is \(2.2 \times 10^{17} \mathrm{\ cm}\), \Alert{and the projected diameter of the cloud is \(\sim 2.8\times 10^{17} \mathrm{\ cm}\). Assuming an approximately spherical shape then gives a total cloud mass of (\(280~M\sun\)) \citep{mcm93}.}}

The close proximity of Orion-S to the Trapezium means there must be an H$^{+}$ region on the side of Orion-S facing the Trapezium. Optical emission lines suggest that it has a density of \(\sim 6000~\mathrm{cm^{-3}}\) \Will{\citep{pog92,gar07}}, corresponding to a gas pressure of \(P/k = 2nT = 1.08 \times 10^{8} \mathrm{\ cm^{-3}\ K}\), assuming a gas temperature of 8500~K\@. \Nick{For the assumed thickness of Orion-S, and the \Will{above estimate of} \(N(\mathrm{H})\), the density deep in the molecular cloud of Orion-S is \(3\times 10^{7} \mathrm{\ cm^{-3}}\). If the environment \Will{were} in \Will{gas-}pressure equilibrium, then the corresponding molecular cloud pressure, \(0.5nT = 2 \times 10^{7}T~\mathrm{cm^{-3}\ K}\) \Will{would correspond} to a temperature of \(7~\mathrm{K}\)\@.} \Will{However, as noted above, this is inconsistent with the presence of gas-phase \form{}, indicating that the gas pressure in the molecular gas must be several times higher than that in the ionized gas.}

\Alert{There are at least four different factors that may account for this \Final{overpressure of the molecular gas}: 
\newcounter{mysave}
  \begin{enumerate}
  \item If the ionization front on the surface of the cloud is D-critical, then the transonic ionized flow from the front contributes an additional ram pressure term, which is roughly equal to the thermal pressure. On the neutral/molecular side, the net flow towards the front is highly subsonic, so the ram pressure term there is negligible. In a steady-state equilibrium, this gives a molecular gas pressure twice that of the ionized gas.
  \item Absorption of the momentum of the stellar photons produces an outwardly directed effective gravity. In a steady-state configuration, in which the molecular gas is forced to remain at rest with respect to the star, this must be balanced by a pressure gradient between the optically thin and optically thick portions of the cloud.
  \item If the cloud is self-gravitating and hydrostatic, then the deeper gravitational potential in the center of the cloud will produce a higher pressure for the molecular gas there than for the ionized gas at the cloud surface.
  \item If the cloud is a transient phenomenon, formed from the dynamic interaction of converging flows, then there is no reason to expect it to be in pressure equilibrium at all.
    \setcounter{mysave}{\value{enumi}}
  \end{enumerate}
  On the other hand, there are are two factors that would tend to work in the other direction:
  \begin{enumerate}
    \setcounter{enumi}{\value{mysave}}
  \item Magnetic fields tend to be relatively weak in ionized nebulae \citep{1981ApJ...247L..77H}, with magnetic pressure smaller than the gas pressure. In molecular clouds, on the other hand the reverse is frequently true, with the magnetic pressure typically exceeding the gas pressure by a factor of 10 or more \citep{1999ApJ...520..706C}.
  \item Turbulent internal motions in molecular clouds are frequently supersonic, with turbulent ram pressure approximately equal to the magnetic pressure. 
  \end{enumerate}
  Therefore, the increase in gas pressure, when going from ionized to molecular gas, indicates that one or more of the first four factors must outweigh factors five and six. 
}

\subsubsection{\Alert{An Equilibrium Cloudy model of Orion-S}}

We perform calculations with version 8.0 of the spectral synthesis code Cloudy, described by \citet{fer78}. Extensions to PDR physics and chemistry are described by \Will{\citet{abe05,abe08}}. The gas-phase composition and dust properties of the Orion environment are well known and are given in \citet{abe04a}.

Our choice of ionizing continuum is characteristic of the dominant source of ionizing photons, the \Alert{O7} star \thC{}. We assume a Kurucz stellar atmosphere with a temperature of 39,600~K, and \(Q(H) = 1.7
\times 10^{49}~\mathrm{photons\ s^{-1}}\). \Will{We include X-ray emission from the stellar wind (see \S~3.1), modeled as a} \(10^{6}~\mathrm{K}\) bremsstrahlung \Will{component}, although this does not affect the results presented below.

\Will{For simplicity, we consider a time-independent, magnetostatic equilbrium for the Orion-S cloud, as has been recently employed to} account for the structure and observed magnetic field in M17 \citep{pel07}.   
\Alert{The current version of Cloudy does not treat self-gravity, and we also ignore gas motions and time-dependent effects, so the model cannot directly address factors 1, 2, or 4, listed in the previous section.}
\Alert{In the model, the total pressure increases as one proceeds into the cloud in such a way that the local pressure gradient exactly counterbalances the acceleration due to absorption of radiation. The total pressure is the sum of the thermal, magnetic, turbulent, and trapped resonance line radiation contributions.} 
For simplicity we assume that the turbulent pressure is in equipartition with the magnetic field and that the magnetic field is \Will{tangled on small scales and} well coupled to the gas, \Alert{so that the magnetic stress tensor reduces to an isotropic pressure that is polytropic, with index \(\gamma_\mathrm{m} =4/3\) \citep{hen05}.} 
The magnetic field of the entire cloud is then set by the field at the illuminated face, \Will{increasing as \(B \sim \rho^{2/3}\)}. The magnetic field in Orion-S is not known but fields are important elsewhere in the Orion environment \citep{abe04a}. The initial magnetic field was taken as \(3~\mu\mathrm{G}\), \Will{which rises to \(850~\mathrm{\mu G}\) deep inside the cloud, where the combined magnetic and turbulent pressures become roughly equal to the thermal pressure.} \Alert{We found that models with stronger magnetic fields were not capable of reproducing the \Final{observed molecular hydrogen density deep in the cloud of \(n \simeq 3 \times 10^7~\pcc\)}, since factors 5 and 6 of \S~\ref{sec:observ-constr} were too effective in counteracting factor 3.}

We assume a closed geometry, in which all radiation must eventually escape in the direction away from the central star cluster, \Will{which is appropriate} if optically thick gas fully surrounds the cluster. Under this assumption, \Will{inwardly directed} diffuse emission is compensated by symmetric emission from \Will{the other side of the \Gary{star cluster}}.  Light can undergo multiple scatterings as it diffuses outward though the cloud, \Alert{so that the total rate of momentum transfer exceeds the single scattering limit, \(L_*/c\), by a factor of a few.} We use the \citet{mcm93} deduced hydrogen column density to set the total thickness of the cloud.

The entire structure and properties of the cloud follows from these assumptions. Cloudy computes the kinetic temperature, ionization, chemistry, optical depths, and emission properties of the gas at each point. The density and radiation field striking the illuminated face are set by observations and the gas pressure at this point is computed. The calculation follows the starlight as it penetrates the cloud. The transmitted continuum sets the pressure and physical conditions. The magnetic field follows from our flux-freezing assumption. The total pressure is set by the assumption of magnetostatic equilibrium. The calculation stops when the observed column density is reached.


\begin{figure*}
  \centering
  \includegraphics[width=\linewidth]{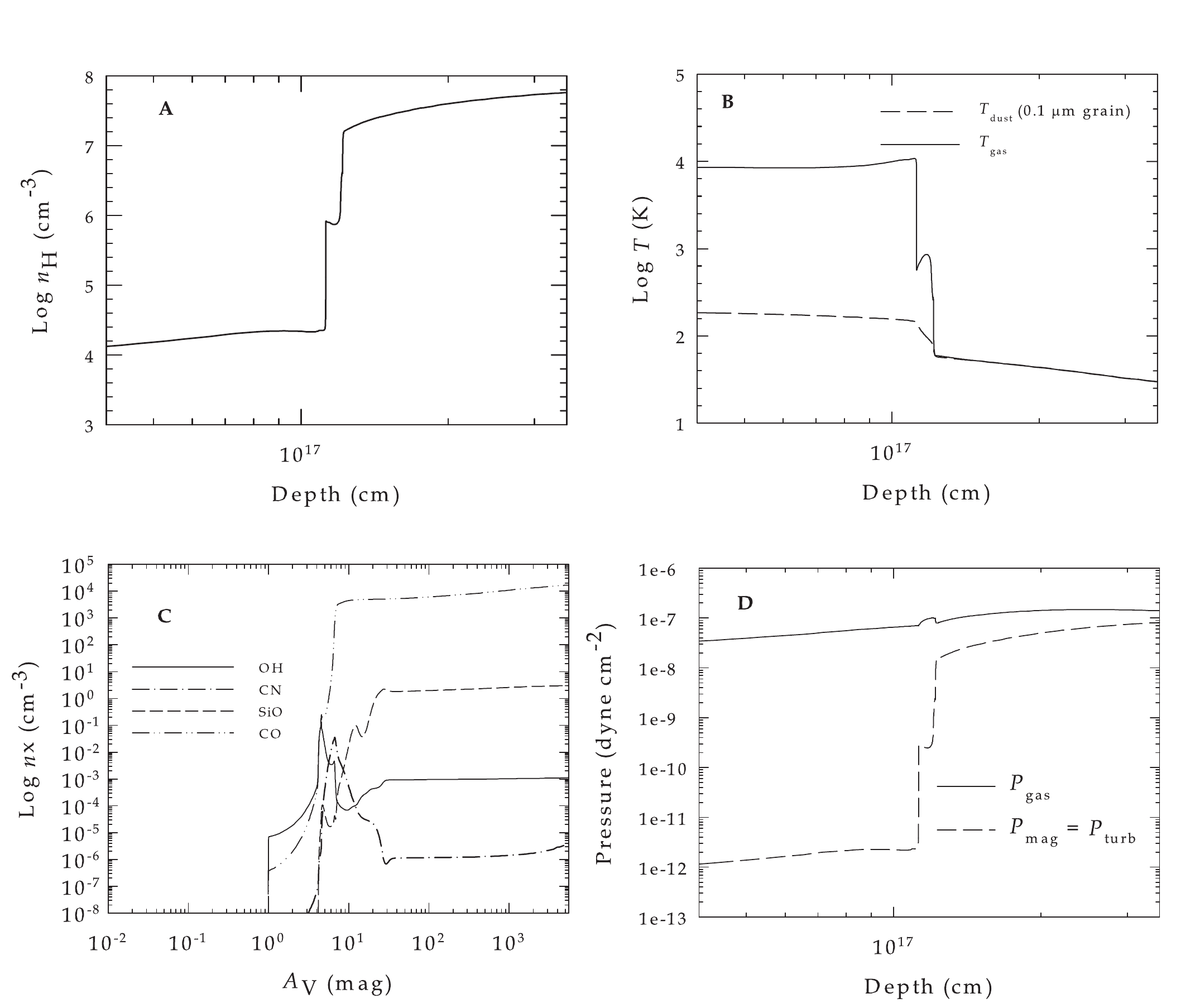}
  \caption{Physical structure for our constant pressure model of the
ionized, atomic, and molecular gas in the Orion-S cloud.  Plotted as a
function of depth into the model are the  (a) density, (b) gas and
dust temperatures (c), molecular abundances, and (d) pressure
constituents.}
  \label{fig:Orion-S-plots}
\end{figure*}
Figure~\ref{fig:Orion-S-plots} shows the results of our calculations \Will{for} the density, temperature, molecular abundances, and pressure as a function of depth into the cloud. The first \Will{criterion} we demand our model reproduce \Will{is clearly satisfied:} that the model yield a temperature high enough to allow H$_{2}$CO to evaporate off of grain surfaces and remain in the gas phase. \Will{The dust temperature (Figure~\ref{fig:Orion-S-plots}\textit{b}) remains above 38~K throughout the model, which, as discussed above in \S~\ref{sec:observ-constr}, is high enough} to inhibit freeze-out of H$_{2}$CO\@. \Will{This is achieved in our model \Gary{purely through external heating by starlight from the Trapezium.}} 

Although we can explain the presence of \Gary{gas-phase} H$_{2}$CO using our model, the chemistry of Orion-S is extremely complex.  At our molecular cloud temperature of 38~K, \Will{some} molecules, such as SiO, CS, and SO\(_{2}\), will freeze out onto grains. Figure~\ref{fig:Orion-S-plots}\textit{c} shows the predicted CO, CN, OH, and SiO abundance as a function of \(A_{V}\).  Our prediction of SiO, once the full effects of freeze out and molecular evaporation off grain surfaces is included, will likely be quite different.  \Final{The current version of Cloudy only treats freeze-out and evaporation processes involving CO, OH, and H\(_2\)O, with the goal of making accurate predictions of gas-phase oxygen deep in molecular clouds.  The freeze-out of other atoms/molecules, such as SiO, is not considered by our models, and neither are grain-surface reactions, other than those that form H\(_2\).  So, while our current analysis is sufficient to deduce that \form{} remains in the gas phase, a full treatment of gas-grain and time-dependent chemical effects is necessary to fully understand all the rich chemical environment of Orion-S.} $\rm ^{12}$CO and $\rm ^{13}$CO lines are optically thick and not sensitive to gas properties such as pressure or column density.  C$\rm ^ {17}$O and C$\rm ^{18}$O should
be optically thin and therefore potentially a good diagnostic of
physical properties.  However, as Orion-S and OMC-1 lie along the same
sightline with similar kinematic properties, the optically thin CO
emission from OMC-1 and Orion-S is difficult to disentangle.
Therefore, we do not present
CO comparisons because of these complexities.

While our current analysis is sufficient to deduce H$_{2}$CO remains in the gas phase, a full treatment of gas-grain and time-dependent chemical effects is necessary to fully understand the chemical environment of Orion-S\@. Such a treatment, while beyond the scope of this work and the current version of Cloudy (Cloudy currently only treats freeze-out of CO, OH, and H\(_{2}\)O), will be the subject of a future investigation.

One appealing aspect of our model is how \Will{we succeed in reproducing} the projected dimensions of Orion-S \Will{under the} assumption of \Alert{magnetostatic equilibrium, with an effective gravity provided by the momentum of the absorbed starlight.}  The density (Figure~\ref{fig:Orion-S-plots}\textit{a}) varies by \(\sim 4~\mathrm{dex}\) over the physical extent of our model, yet our calculation reaches \(N(\mathrm{H}) \sim 4 \times 10^{24}~\mathrm{cm^{-2}}\) at a depth of \(2.4\times 10^{17}~\mathrm{cm}\), or within 20\% of the \Will{observed} projected thickness. \Alert{Despite this success, there are \Final{arguably} some \Final{deficiencies} in our model assumptions, of which the most serious are the neglect of self-gravity and the \Final{rather} weak magnetic field. Using the observed mass and size of the cloud (\S~\ref{sec:observ-constr}), one finds \Nick{the escape speed at the cloud surface to be \(7.3~\kms\), meaning that the molecular gas is gravitationally bound, while the ionized gas is gravitationally unbound.} Thus, factor~3 of \S~\ref{sec:observ-constr} is likely to contribute significantly to the overpressure of the molecular gas, so that a self-gravitating model would allow the magnetic field in the cloud to be significantly higher, while still satisfying the observational constraints. However, it is not clear that any equilibrium self-gravitating model is possible. Unless the turbulent and magnetic pressures are more than 100 times greater than given by our model, the cloud mass is higher than both the Jeans critical value and the magnetic critical value, and so the cloud may be dynamically collapsing on a free-fall timescale of a few times \(10^4\)~years.} 

\Final{In summary, our equilibrium Cloudy model represents a necessary compromise, including as it does a detailed simulation of the relevant microphysics, but at the expense of a simplified treatment of some of the relevant macrophysics. As such, although it is broadly successful in accounting for the observed properties of the Orion-S cloud using almost no free parameters, it should not be considered the last word in modeling this region, especially in the light of the uncertainties outlined in the previous paragraph.}


\subsection{What is the Impact of a Putative High Relative Velocity of \ori?}
\label{sec:what-impact-putative}

In \S\ 2.7 we saw that \ori\ appears to be moving away from the OMC at about 13 \kms.  Since this star is only about 0.2 pc in front of the local MIF, \ori\ would have been at the present location of the MIF only 15,000 years ago. This is simply a reference number for the dynamic timescale as one would expect that \ori\ would have continuously ionized the gas immediately surrounding it, thus producing a constantly changing structure. It does appear that \ori\ has only recently entered a region of low density gas and has just begun to photo-ionize a much larger volume of space (the Huygens region and the EON). This is a dramatically new way of viewing the Orion Nebula and means that the object need not be in quasi dynamic equilibrium. A short illumination age of the ONC stars is quite attractive because the well determined mass loss rates of the proplyds close to \ori\ indicate that they should have been destroyed through photo-ablation within the last 1.5x10$^{4}$--1.1x10$^{6}$ years \citep{chu87,hen02,smi05,ode08a}. The absence of any evidence for destruction indicates that their illumination ages by \ori\ are no more than a few times 1.5x10$^{4}$ years.

If one accepts the scenario that \thC{} has emerged from a much higher
density environment during the last 10,000~years, then it is worth
considering how this picture would change the expected ionization,
density, and dynamics of the present-day nebula. If the surrounding
density was much higher in the past, then the \hii{} region would have
been much smaller. For example, in an ambient density of \(n =
10^6~\pcc\), the Str\"omgren radius would be only
\(0.007~\mathrm{pc}\), or 3.2\arcsec{}. As \thC{} left the
high-density environment, an R-type ionization front would propagate
into the low-density gas at a speed of \(\Phi_\mathrm{H}/n \ge
1000~\kms\), \Will{where \(\Phi_\mathrm{H}\) is the ionizing flux,
  defined in \S~\ref{sec:phys-cond-oiii} above.} This speed is much
larger than the stellar velocity, so that deviations from static
photoionization equilibrium would be fleeting and difficult to
detect. Assuming a density \(n = 100~\pcc\) for the low-density gas,
the ionization front would reach its equilibrium Str\"omgren value
(\(\simeq 3\)~pc) in only 1000~years. The maximum recombination front
velocity is also of order \(1000~\kms\) so that the gas would
recombine effectively instantaneously in the high-density regions
vacated by the star. The timescale for molecular hydrogen formation on
dust grains is approximately \(1000~(10^6~\pcc/n)~\mathrm{yr}\)
\citep{tie85}, so that, unless the previously ionized gas was of
rather low density, there has been time for it to return to a fully
molecular state.

The density structure of the ionized gas is intimately tied to its dynamics, with typical internal velocities of order the ionized sound speed, which is very similar to the putative stellar velocity of \(13~\kms{}\). It seems likely, therefore, that the ionized density structure will not be greatly affected by the stellar
motion. Numerical simulations confirm \citep{2005ApJ...627..813H,2006ApJS..165..283A} that an approximately steady-state champagne flow is established on a timescale equal to the sound crossing
time. It is therefore at the largest scales in the nebula (\(>1~\mathrm{pc}\)) that one would expect to find the clearest evidence for non-equilibrium dynamics induced by the stellar motion. However, uncertainties in the exact ambient density distribution and the effect of the stellar wind would make it
difficult to draw clear conclusions. Another possible observational diagnostic of a moving ionized star would be the dense neutral wake, which numerical simulations predict should form behind the star
\citep{1997RMxAA..33...73R, kra07}. Again, observational detection of such a wake would be very challenging, chiefly due to confusion brought on by the rich profusion of structures at all scales in the molecular cloud behind the nebula.

A very young age for the present configuration of the Orion Nebula could also remove the dilemma presented by the fact that the H$^{+}$ layer lying between \ori\ and the Veil is moving towards the Veil with a closure velocity of  about 18 \kms\ and there is no evidence of an interaction with the Veil. \citet{abe06} calculate that the collision time with the Veil is about 40,000 years. Without a detailed model of the expected very time-dependent structure, one cannot assess the importance of this number except that it too indicates that the present configuration is quite young.

In conclusion, it seems impossible to definitively confirm or rule out
the moving star hypothesis, based on existing observational
evidence. The strongest argument in favor remains the short mass-loss
lifetimes derived for the proplyds in the inner nebula. The strongest
arguments against are: (1)~the required coincidence that \thC{}
currently resides so close to the center of the Trapezium cluster,
despite not being dynamically bound to the cluster, and (2)~the
difficulty in explaining how \thC{} acquired such a high velocity. The
mass of the current \thC{} multiple system is \(\simeq 50~M_\odot\)
\citep{kra07}, so assuming that it originally had a velocity typical
of ONC members, then to achieve its current velocity it must have
ejected a \(\simeq 10~M_\odot\) companion at \(65~\kms\), or even
faster for a less massive star. Note that these high stellar velocities mean that many-body interactions at the level of the star cluster are unlikely to play a role since the velocity dispersion of the cluster is only a      few \kms (Jones \&\ Walker 1988) and disruption of binaries by the cluster potential has been shown to be ineffective for binary separations \(< 0.5''\) (Reipurth et al. 2007).

\Will{One further possibility that bears examination is that the
  velocity kick of the ionizing star was acquired much more
  recently. This would sidestep most of the arguments discussed above,
  both for and against the moving star hypothesis. For example,
  \citet{tan04, 2008arXiv0807.3771T} has proposed that the runaway B
  star known as the Becklin-Neugebauer object
  \citep{1967ApJ...147..799B} was ejected from the \thC{} multiple
  system 4500~years ago (but see \citealt{gom05, gom08} for an
  alternative viewpoint). If the anomalous velocity of \thC{} is due
  to this event, then it will only have moved about \(0.05\)~pc since
  then, which is very small compared with the size of the nebula, so
  that observable consequences are unlikely. However, one serious
  problem with this proposal is that the measured radial velocity of
  BN is \(+13~\kms{}\) with respect to the ONC rest frame
  \citep{1983ApJ...275..201S}, which is the same absolute value as
  that recently determined for \thC{} \citep{sta08}, but in the
  opposite direction. BN's mid-IR luminosity of \(2600~L_\sun\)
  (\citealp{1998ApJ...509..283G}, re-adjusted to a distance of 440~pc)
  is consistent with a dust-enshrouded main-sequence B2 to B3 star,
  implying a mass of \(9 \pm 1~M_\sun\), which is more than 5 times
  less than the mass of \thC{}. Therefore, as discussed in the
  previous paragraph, momentum conservation means that BN ought to
  have a much higher redshifted velocity than is observed if it has
  truly been ejected from the \thC{} system in the recent past.}

\subsection{Summary and Conclusions}

\Bob{In this paper we have been able to establish several remarkable new properties of the Orion Nebula.

1. There is an incomplete shell of high ionization gas shaped by the high velocity wind arising from \ori\ and open in the direction of detected x-ray emission.

2. The highest surface brightness portion of the Orion Nebula occurs to the southwest of the Trapezium stars, in the region called Orion-S, a center of star formation and stellar outflow.

3. The Orion-S region is caused by an optically thick molecular cloud which lies within the ionized cavity of the Orion Nebula. This Orion-S Cloud is the host of a separate region of star formation and is sufficiently cold in the middle that it can produce \form\ lines in absorption.

4. It is possible that \ori\ has a large radial velocity with respect to the ONC and other members of the Trapezium. The observations of the nebula do not contradict such a possibility and the presence of a large radial velocity would resolve the conundrum of the lack of destruction of the disks in proplyds close to \ori\ and would explain why the inner H$^{+}$ layer expanding towards the observer has not yet reached the Veil feature.}

\acknowledgments

CRO's work was supported in part by the Space Telescope Science
Institute grants GO 10921 and GO 10967. NPA acknowledges financial
support through the National Science Foundation under Grant
No. 0094050, 0607497 to the University of Cincinnati. GJF thanks the
NSF (AST 0607028) and NASA (NNG05GD81G) for support. \Will{WJH and SJA
  acknowledge financial support from DGAPA-UNAM, Mexico (PAPIIT
  IN110108 and IN100309).}



{\it Facilities:} \facility{HST (WFPC2)}.

\end{document}